\documentclass[journal]{IEEEtran}   

\ifCLASSINFOpdf
\else
\fi

\newcommand{\dd}{\mathrm{d}}
\newcommand{\zerospace}{\setlength{\arraycolsep}{0pt}}

\usepackage{amsmath}
\usepackage{graphicx,psfrag} 
\usepackage{amsmath,amsthm,amsfonts,amssymb,bm} 
\usepackage[ruled]{algorithm2e}
\usepackage{algpseudocode}
\usepackage{xcolor}
\usepackage{bbm}             
\usepackage{tabularx}        

\newcommand{\bs}{\boldsymbol}

\newtheorem{proposal}{Proposal}

\begin{document}

\title{
Estimation for High-Dimensional Multi-Layer Generalized Linear Model -- Part~II: The ML-GAMP Estimator
}

\author{
{Qiuyun Zou},
{Haochuan Zhang*},
and
{Hongwen Yang}
\thanks{
Q. Zou was with School of Automation, Guangdong University of Technology, Guangzhou 510006, China, and is now with School of Information and Communication Engineering, Beijing University of Posts and Telecommunications, Beijing 100876, China (qiuyunzou@qq.com)
}
\thanks{
H. Zhang is with School of Automation, Guangdong University of Technology, Guangzhou 510006, China (haochuan.zhang@gdut.edu.cn).
}
\thanks{
H. Yang is with School of Information and Communication Engineering, Beijing University of Posts and Telecommunications, Beijing 100876, China (yanghong@bupt.edu.cn).
}
\thanks{
*Corresponding author: H.~Zhang.
}

}

\maketitle

\begin{abstract}
This is Part II of a two-part work on the estimation for a multi-layer generalized linear model (ML-GLM) in large system limits. In Part I, we had analyzed the asymptotic performance of an exact MMSE estimator, and obtained a set of coupled equations that could characterize its MSE performance. To work around the implementation difficulty of the exact estimator, this paper continues to propose an approximate solution, ML-GAMP, which could be derived by blending a moment-matching projection into the Gaussian approximated loopy belief propagation. The ML-GAMP estimator is then shown to enjoy a great simplicity in its implementation, where its per-iteration complexity is as low as GAMP. Further analysis on its asymptotic performance also reveals that, in large system limits, its dynamical MSE behavior is fully characterized by a set of simple one-dimensional iterating equations, termed state evolution (SE). Interestingly, this SE of ML-GAMP share exactly the same fixed points with an exact MMSE estimator whose fixed points were obtained in Part I via a replica analysis. Given the Bayes-optimality of the exact implementation, this proposed estimator (if converged) is optimal in the MSE sense.
\end{abstract}


\IEEEpeerreviewmaketitle

\section{Introduction}
\subsection{Problem Statement and Recap on Part I}
\begin{figure}[!t]
\centering
\includegraphics[width=0.5\textwidth]{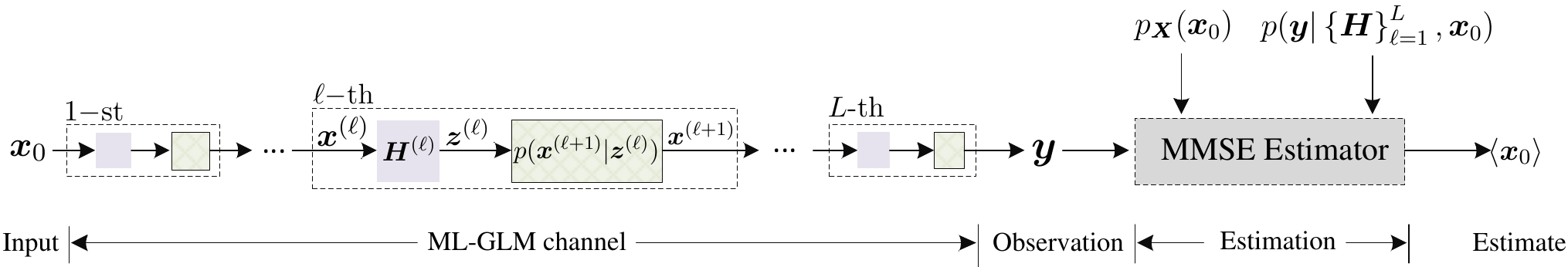}
\caption{System model of estimation in ML-GLM: random input $\to$ ML-GLM network $\to$ observation $\to$ MMSE estimator $\to$ output estimate.
}
\label{fig:mlglm}
\end{figure}

This two-part work considers the estimation of high-dimensional random signal in a multi-layer generalized linear model (ML-GLM), which is illustrated as Fig.~\ref{fig:mlglm}. In the figure, $\bs{x}_0$ denotes the initial random input, whose distribution is factorable and known perfectly by the estimator, i.e.,
$
\bs{x}_0 \sim \mathcal{P}_{X}(\bs{x}_0)
=
    \prod_{i=1}^{N_1} \mathcal{P}_X (x_{0i})
,
$ 
$\bs{y}$ denotes the observation attained from the ML-GLM network  of $L$ layers, and $\langle \bs{x} \rangle$ is the MMSE estimator's output, either exact or approximate. Particularly, the $\ell$-th layer expands as ($1\leq \ell \leq L$)
\begin{equation}
\!\to\!\!
\bs{x}^{(\ell)}
\!\!\to\!\!
\boxed{ 
\bs{H}^{(\ell)}\bs{x}^{(\ell)}}
\!\!\to\!\!
\bs{z}^{(\ell)}
\!\!\to\!\!
\boxed {\mathcal{P}(\bs{x}^{(\ell+1)}|\bs{z}^{(\ell)})}
\!\!\to\! \bs{x}^{(\ell+1)}
\!\!\to\!\!
\label{A1}
\end{equation}
where $\bs{x}^{(\ell)}\in \mathbb{R}^{N_{\ell}}$ is its input, and $\bs{H}^{(\ell)}\in \mathbb{R}^{N_{\ell+1}\times N_{\ell}}$ is a deterministic weighting matrix that linearly mixes up the input to yield $\bs{z}\in \mathbb{R}^{N_{\ell+1}}$. This weighted result $\bs{z}$ is then activaed by a random mapping, whose transitional/conditional probability density function (p.d.f.) is also factorable:
$
\mathcal{P}(\bs{x}^{(\ell+1)}|\bs{z}^{(\ell)})
=
    \prod_{a=1}^{N_{\ell+1}}\mathcal{P}(x_a^{(\ell+1)}|z_a^{(\ell)})
.
$ 
The weighting matrix above is known perfectly to the estimator, and in each experiment, the elements of this matrix are drawn independently from the same Gaussian ensemble of zero mean and $1/N_{\ell+1}$ variance (to ensure a unit row norm).
To matain notational consistency, we also initialize: $\bs{x}^{(1)} := \bs{x}_0$, and $\bs{x}^{(L+1)} := \bs{y}$.
Since we consider exclusively the limiting performance of the MMSE estimators, the following assumptions are made throughout the paper: $N_{\ell}\to \infty$, while ${N_{\ell+1}}/{N_{\ell}} \to \alpha_{\ell}$, i.e., all weighting matrices are sufficiently large in size, but the ratios of their row numbers to culumn numbers are fixed and bounded.

The target of an exact MMSE estimator is to generate an estimate $\langle x_k\rangle$ for every input element $x_{0k}$ using ($k=1,\cdots,N_1$)
\begin{equation}
\langle x_k\rangle
=
\arg \min_{\hat{x}_k} \mathbb{E} \left[ \|\hat{x}_k-x_k\|^2 \right]
=\mathbb{E}\left[x_k \left|\bs{y},\{\bs{H}^{(\ell)}\} \right. \right]
\label{eq:MMSE_est}
\end{equation}
where the last expectation is taken over a marginal posterior
\begin{align}
\mathcal{P}(x_{0k}|\bs{y},\{\bs{H}^{(\ell)}\})=\int \mathcal{P}(\bs{x}_0|\bs{y},\{\bs{H}^{(\ell)}\})\dd \bs{x}_{0\backslash k}
,
\end{align}
whose integration is $(N_1-1)$-fold, $\bs{x}_{0\backslash k}$ equals $\bs{x}_0$ except its $k$-th element moved, and the joint p.d.f. $\mathcal{P}(\bs{x}_0|\bs{y},\{\bs{H}^{(\ell)}\})$ is:
\begin{align*}
\mathcal{P}(\bs{x}_0|\bs{y},\{\bs{H}^{(\ell)}\})
=
    \frac{
        \mathcal{P}_{X}(\bs{x}_0)\mathcal{P}(\bs{y}|\bs{x}_0,\{\bs{H}^{(\ell)}\})
    }
    {\int \mathcal{P}_{X}(\bs{x}_0)\mathcal{P}(\bs{y}|\bs{x}_0,\left\{\bs{H}^{(\ell)}\right\})\dd \bs{x}_0
    }
.
\end{align*}
Due to the difficulty in the evaluation of the $(N_1-1)$-fold integrals above, an exact implementation of the MMSE estimator is infeasibility \cite{bishop2006pattern}, and as a consequence, this Part II paper considers its effective approximation, and proposes a new estimator called multi-layer generalized approximate message passing (ML-GAMP), which is later shown  MSE-optimal and computationally efficient.

\begin{figure}[!t]
\centering
\includegraphics[width=0.45\textwidth]{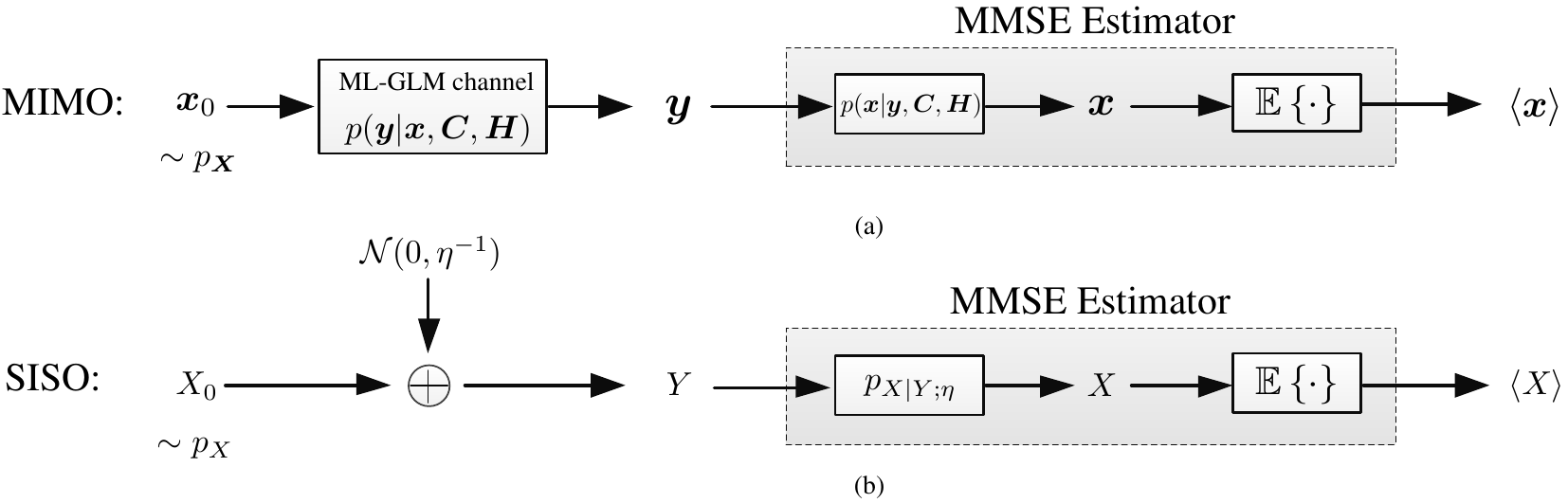}
\caption{The MIMO-to-SISO decoupling principle
}
\label{fig:RroChannel}
\end{figure}

But before proceeding further, a short recap on Part I's findings from replica analysis is given first, considering exclusively an exact MMSE estimator:

1) In terms of joint input-and-estimate distribution, the original estimation problem of MIMO nature was identical to that of a simple SISO estimation problem facing no self-interference (caused by the linear weighting), no nonlinear distortion, (caused by the random mapping), but only an effective AWGN. That is,
\begin{align}
\ (x_{0k},\langle x_{k}\rangle)
& \doteq
(X_0,\langle X \rangle)
,
\quad \forall k,
\end{align}
As an illustration, Fig. \ref{fig:RroChannel} depicts the MIMO-to-SISO decoupling principle for a relatively simple case of two layers.

2) The noise level of this AWGN could be further determined by solving a set of coupled equations (given in Claim 1 of Part I), whose dependency on the linear weighting and the random mapping was given explicitly (also in Claim 1).

3) As a byproduct of the replica analysis, the average MSE of the exact MMSE estimator could be computed directly from the fixed-point results, in which case time-consuming simulations in the Mote Carlo manner are no longer needed.

\subsection{Contributions of Part II}
Given the infeasibility of an exact MMSE implementation in large system limit, we propose in Part II (this paper) a new estimator called ML-GAMP, which is asymptotically optimal in MSE sense but still enjoys a great computational simplicity. For its derivation, we, instead of approximating directly the exact MMSE estimation, resort to the message passing in random graphical models, where a method developed by us in an earlier study \cite{zou2018concise} could be applied to find new algorithms for the posterior mean estimation (PME). The method developed was in general under the framework of the celebrated loopy belief propagation (LBP) \cite{Murphy-arxiv13-LBP}, but some of its message updating had blended the traditional LBP with another powerful variational inference tool, the expectation propagation (EP) \cite{minka2001family, bishop2006pattern}, or the expectation consistent (EC) \cite{opper2005expectation}. This blended method was shown \cite{zou2018concise} capable of recovering the GAMP algorithm \cite{Rangan-arxiv10-GAMP, rangan2011generalized} in GLM (generalized linear model), and will continue to be demonstrated applicable to the more general ML-GLM setting.

In what follows, we summarize the contributions of Part II:

1)
A new estimator called ML-GAMP is proposed for the estimation of the ML-GLM problem.
Comparing to prior works that are closely related, our proposed estimator provides a better tradeoff between performance and complexity. To be specific, although both could attain an MSE-optimal performance, our estimator converges in a speed much faster than the ML-AMP \cite{manoel2017multi} proposed by Manoel \emph{et al.}, which we believe was the first to put forward a solution for the challenging problem of ML-GLM estimation.
An investigation to this difference in convergence rate is also carried out, which suggests that our iteration could make full use of the messages most recently updated while the ML-AMP \cite{manoel2017multi} put more faith in the outdated ones, so the two diverges and ours becomes superior (especially in the correct-updating-dominant cases).
Another solution that solved the ML-GLM problem was ML-VAMP \cite{fletcher2018inference} proposed by Fletcher \emph{et al.}, which was a significant extension to VAMP \cite{rangan2019vector}, developed by Rangan \emph{et al.} and tightly connected with EP \cite{minka2001family} and EC \cite{opper2005expectation}. This ML-VAMP \cite{fletcher2018inference}, however,  required during its iterations (at least in the very first) to perform a singular value decomposition (SVD) of $O(N^3)$ complexity on each weighting matrix, which inevitably compromised the computational efficiency, as opposed to the scalar-only operations in ML-AMP \cite{manoel2017multi} and ours. In this context, the proposed estimator has less computational burden than the ML-VAMP \cite{fletcher2018inference}.

2)
An set of simple one-dimensional iterative equations, termed the state evolution (SE), is derived from ML-GAMP, which could characterize fully the proposed estimator's dynamical MSE behavior under a mild assumption on weighting matrices of zero-mean and bouned-variance Gaussian distributions.  Given the SE, the average MSE of ML-GAMP could be easily obtained by iterating the SE. The accuracy of this SE iteration is then validated via numerical experiments, which clearly show that not only the overall MSE at the final step of the estimation is accurate, but also those in the middle (algorithm not-yet converged) steps are accurately captured by the SE.

3)
A perfect agreement is observed between the above SE and the fixed point equations of an exact MMSE estimator (given in Claim 1 of Part I). Since the fixed point equations were analyzed from an exact (rather than approximate) MMSE implementation and the average MSE of the exact MMSE estimator could be computed directly from the fixed equations, the above agreement provides an interesting and very important piece of evidence that the proposed ML-GAMP (if converged) could attain the same MSE performance as an exact MMSE estimator, and thus is Bayes-optimal in the MSE sense.

\section{ML-GAMP: A Message Passing Algorithm} \label{sec:algo}
\begin{algorithm}[!t]
\caption{The Proposed ML-GLM Estimator}
\label{alg:CML-GAMP}
\KwIn{observed signal $\bs{y}$, measure matrix $\bs{H}^{(\ell)}$, transition distribution $\mathcal{P}(\bs{x}^{(\ell+1)}|\bs{z}^{(\ell)})$ for $(\ell=1,\cdots, L)$ and prior distribution $\mathcal{P}(\bs{x})$.}
\textbf{Initiation}: $t=1$, $Z_a^{(\ell)}(1)=\prod_{k=1}^{\ell}\alpha_k$, $V_a^{(\ell)}(1)=1$ for $(\ell=1,\cdots,L)$ and $(a=1\cdots, N_{\ell+1})$.

\KwOut{Estimator $\hat{\bs{m}}^{(1)}$}
\For{$t=1,\cdots,T$ }
{
   \For{$\ell=L,\cdots,1$ }
   {
      \begin{subequations}
      \begin{align}
      \tilde{z}_a^{(\ell)}(t)
      &=\mathbb{E}_{\tilde{\zeta}_a^{(\ell)}(t)}\left[\tilde{\zeta}_a^{(\ell)}(t)\right]
      \label{AD1}\\
      \tilde{v}_a^{(\ell)}(t)
      &=\text{Var}_{\tilde{\zeta}_a^{(\ell)}(t)}\left[\tilde{\zeta}_a^{(\ell)}(t)\right]
      \label{AD2}\\
      s_a^{(\ell)}(t)&={(\tilde{z}_a^{(\ell)}(t)-Z_a^{(\ell)}(t))}/{V_a^{(\ell)}(t)}
      \label{AD3}\\
      \tau_a^{(\ell)}(t)&=(V_a^{(\ell)}(t)-\tilde{v}_a^{(\ell)}(t))/{(V_a^{(\ell)}(t))^2}
      \label{AD4}\\
      \Sigma_i^{(\ell)}(t)&=\left[{\sum\nolimits_a|H_{ai}^{(\ell)}|^2\tau_a^{(\ell)}(t)}\right]^{-1}
      \label{AD5}\\
      R_i^{(\ell)}(t)&=\hat{m}_i^{(\ell)}(t)+\Sigma_i^{(\ell)}(t)\sum\nolimits_a ({H_{ai}^{(\ell)}})^{*}s_a^{(\ell)}(t)
      \label{AD6}
      \end{align}
      \end{subequations}
   }
   \For{$\ell=1,\cdots,L$}
   {
    \begin{subequations}
    \begin{align}
    \hat{m}_i^{(\ell)}(t+1)&=\mathbb{E}_{\xi_i^{(\ell)}(t+1)}\left[\xi_{i}^{(\ell)}(t+1)\right]
    \label{AD7}\\
    \hat{v}_i^{(\ell)}(t+1)&=\text{Var}_{\xi_i^{(\ell)}(t+1)}\left[\xi_{i}^{(\ell)}(t+1)\right]
    \label{AD8}\\
    V_a^{(\ell)}(t+1)&=\sum\nolimits_i |H_{ai}^{(\ell)}|^2\hat{v}_i^{(\ell)}(t+1)
    \label{AD9}\\
    Z_a^{(\ell)}(t+1)
    &\!=\!
    \sum\nolimits_i \! H_{ai}^{(\ell)}\hat{m}_i^{(\ell)}(t\!+\!\!1)
    \!-\!\!
    V_a^{(\ell)}(t\!+\!\!1)s_a^{(\ell)}(t)
    \label{AD10}
    \end{align}
    \end{subequations}
   }
}
\end{algorithm}

This section provides a description for the proposed ML-GAMP, together with its detailed derivation via a moment-matching-based LBP method, and a simplified version to further improve its computational efficiency.


\subsection{Description of the Algorithm}
\begin{proposal}
To estimate $\bs{x}$ for the ML-GLM problem depicted as \ref{fig:mlglm}, we propose a new algorithm, termed \emph{ML-GAMP}, whose detail is given in Algorithm \ref{alg:CML-GAMP}.
\end{proposal}

On ML-GAMP, we have the following remarks.\\
\underline{Remark 1}:
The ML-GAMP degenerates smoothly to the celebrated AMP \cite{donoho2009message} and GAMP \cite{Rangan-arxiv10-GAMP}. Thanks to the generality of the ML-GLM model, the proposed estimator embraces many existing results as special cases. For instance, by particularizing $L=1$, it recovers GAMP \cite{Rangan-arxiv10-GAMP} which reads
\begin{align*}
\tilde{z}_a(t)&=\mathbb{E}_{\zeta_a(t)}\left\{\zeta_a(t)\right\}\\
\tilde{v}_a(t)&=\text{Var}_{\zeta_a(t)}\left\{\zeta_a(t)\right\}\\
s_a(t)&={(\tilde{z}_a(t)-Z_a(t))}/{V_a(t)}\\
\tau_a(t)&=(V_a(t)-\tilde{v}_a(t))/{(V_a(t))^2}\\
\Sigma_i(t)&=\left[{\sum\nolimits_a|H_{ai}|^2\tau_a(t)}\right]^{-1}\\
R_i(t)&=\hat{m}_i^{(\ell)}(t)+\Sigma_i(t)\sum\nolimits_a ({H_{ai}})^{*}s_a(t)\\
\hat{m}_i(t+1)&=\mathbb{E}_{\xi_i(t+1)}\left\{\xi_{i}(t+1)\right\}\\
\hat{v}_i(t+1)&=\text{Var}_{\xi_i(t+1)}\left\{\xi_{i}(t+1)\right\}\\
V_a(t+1)&=\sum\nolimits_i |H_{ai}|^2\hat{v}_i(t+1)\\
Z_a(t+1)&=\sum\nolimits_iH_{ai}\hat{m}_i(t+1)-V_a(t+1)s_a(t)
\end{align*}
where
\begin{align}
\zeta_a(t)
& \sim
    \frac{\mathcal{P}(y_a|z_a)\mathcal{N}(z_a|Z_a(t),V_a(t))}{\int \mathcal{P}(y_a|z_a)\mathcal{N}(z_a|Z_a(t),V_a(t))\dd z_a}
\\
\xi_{i}(t+1)
& \sim
    \frac{p_X(x_i)\mathcal{N}(x_i|R_i(t),\Sigma_i(t))}{\int p_X(x_i)\mathcal{N}(x_i|R_i(t),\Sigma_i(t)) \dd x_i}
.
\end{align}
Also in the single-layer case, if we continue to particularize the random activation $\mathcal{P}(y_a|z_a)$ as Gaussian, say, $\mathcal{N}(z_a|y_a,\sigma_w^2)$, the ML-GAMP algorithm will further degenerate to AMP \cite{donoho2009message} after substituting the results below into  $\hat{m}_i(t+1)$, $\hat{v}_i(t+1)$, $V_a(t+1)$, $Z_a(t+1)$
\begin{align}
\Sigma_i(t)
    &=\left(\sum_a \frac{|H_{ai}|^2}{\sigma_w^2+V_a^t}\right)^{-1}
\\
R_i(t)
    &=m_i(t)+\Sigma_i(t)\sum_a \frac{H_{ai}^{*}(y_a-Z_a^t)}{\sigma_w^2+V_a^t}
.
\end{align}

\noindent\underline{Remark 2}:  Comparing to the existing results, the proposed algorithm offers a better tradeoff between estimation performance and computational efficiency. As seen from Fig. \ref{fig:sim_MLGAMP&MLAMP} in the experiment section, the proposal noticeably outperforms ML-AMP \cite{manoel2017multi} in the early stage of its iterations. ML-AMP \cite{manoel2017multi} was a competing state-of-the-art estimator proposed by Manoel, Krzakala, M{\'e}zard, and Zdeborov{\'a}, and, to the best of our knowledge, it was the first (in the AMP family, at least) to solve the ML-GLM estimation problem. Comparing to the ML-AMP, our proposal here converges in a much faster speed, and thus, it could offer an estimate far more accurate than ML-AMP \cite{manoel2017multi} in scenarios of stringent time budget, e.g., in interactive video communications, where the maximum number of iterations is strictly limited.
When it comes to computational complexity, the proposed algorithm resembles ML-AMP \cite{manoel2017multi} in a sense that both are having a complexity dominated by some matrix-vector multiplications of $O(N^{(\ell+1)} N^{(\ell)})$ in the worst case, c.f. (\ref{AD5}), (\ref{AD6}), (\ref{AD9}), and (\ref{AD10}). That is, both estimators are as efficient as the celebrated GAMP \cite{Rangan-arxiv10-GAMP}, whose computational simplicity is a key appeal.
Comparing to ML-VAMP \cite{Fletcher-arxiv17-ML-VAMP}, another state-of-the-art estimator developed by Fletcher, Rangan, and Schniter,  the proposed algorithm is computationally more efficient, because it does not rely on singular value decomposition (SVD), an expansive operation of $O(N^3)$ complexity, while in contrast, ML-VAMP \cite{Fletcher-arxiv17-ML-VAMP} required to perform SVD (at least once) for each of its weighting matrices.

\begin{figure}[!t]
\centering
\includegraphics[width=0.5\textwidth]{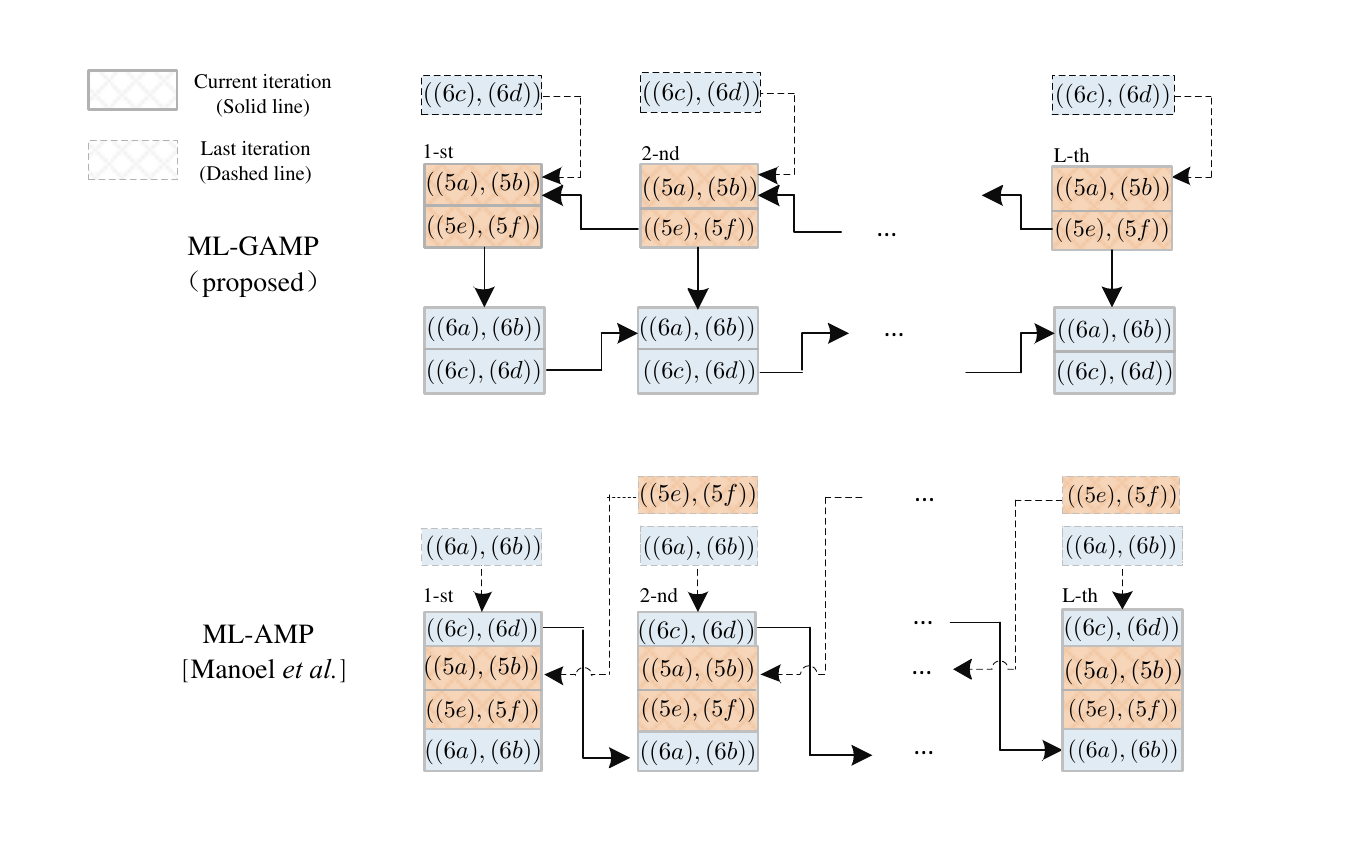}
\caption{ML-GAMP (proposed) Vs. ML-AMP \cite{manoel2017multi}, suggesting that the proposed algorithm puts more faith in the messages updated most recently.
}
\label{fig:IterationOrder}
\end{figure}

\noindent\underline{Remark 3}: To differentiate our proposal further from the existing ML-AMP \cite{manoel2017multi}, we compare the two in more details and find out that the proposed algorithm relies on the most recently updated messages, while ML-AMP \cite{manoel2017multi} put more faith in some older ones. Fig~\ref{fig:IterationOrder} provides a more detailed illustration, upon which we make these comments: \\
$\bullet$
The proposed algorithm updates its messages in a \emph{direction-by-direction} manner, i.e., first backward, then forward, after that backward again, and repeat it until converged, but the ML-AMP \cite{manoel2017multi} adopted a different \emph{layer-by-layer} manner that first updates (both directions in) the 1st layer, and then (both directions in) the 2nd, after that the 3rd, and so on.
\\
$\bullet$ To see how they differ in more detail, we consider a particular block of equations, (6a)-(6d).
In the proposed algorithm (i.e., the upper half of the figure), this block is updated using results from the same current iteration (either from an adjacent layer in the same direction, or from the same layer but in a different direction).
In contrast, this block in the ML-AMP \cite{manoel2017multi} case (i.e., the lower half of the figure) was updated combining results from an adjacent layer in the same iteration and the same layer but in the last iteration.\\
Noticing that the other block, i.e., (5a)-(5f), is updated similarly in both algorithms, one would come to conclude that the proposed algorithm relies more on the recent updates, while the ML-AMP \cite{manoel2017multi} was putting more faith in the earlier ones. It could also be inferred that, in a case dominated by correct estimates, the proposed algorithm may get extra benefit from its use of the more recent messages, and thus exhibit a faster convergence rate. This is indeed evidenced by the simulation results given later in Fig. \ref{fig:sim_MLGAMP&MLAMP} of the paper.

\subsection{Derivation of the Algorithm}

\begin{figure}[!t]
\centering
\includegraphics[width=0.5\textwidth]{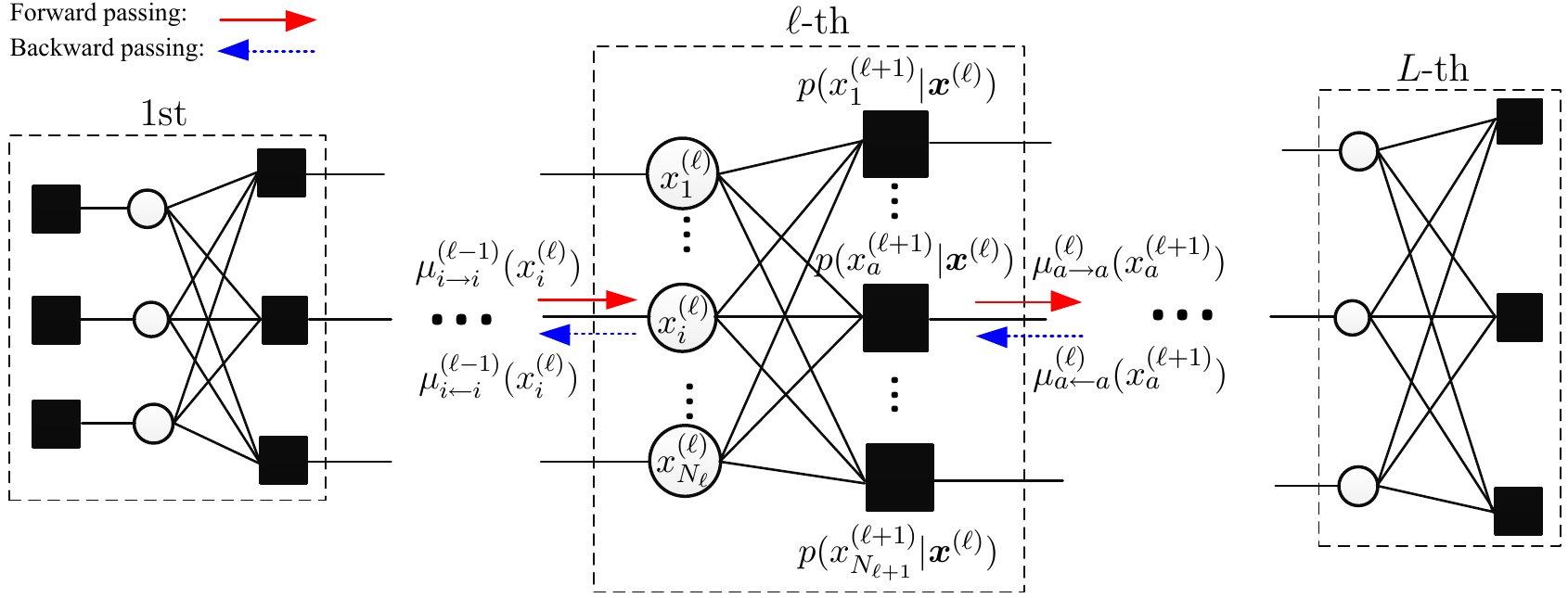}
\caption{Factor graph for the ML-GLM estimation problem
}
\label{fig:FactoGraph}
\end{figure}

To derive ML-GAMP, we follow a method used in our previous paper \cite{zou2018concise} that blended an moment-matching projection into the sum-product loopy belief propagation (LBP) \cite{murphy1999loopy}, whose a factor graph is given in Fig. \ref{fig:FactoGraph}. In the figure, empty circles (variable nodes) stand for the latent and hidden variables to estimate, shaded boxes (factor nodes) represent the functions to which these variables are associated, and an edge connecting the two indicates that these two have an explicit argument-and-function dependency. On the factor graph, a sum-product LBP associates to each edge in either of the two directions a message that is in essence a marginal distribution (either prior or likelihood) of the variable node. In the context of a ML-GLM, we denote the message from one node $m$ to another node $n$ as $\mu_{m \to n}(x_k)$, with $x_k$ being the variable node between the two, and add a superscript $(\ell)$ to index its layer number. We further append an iteration number  $(t)$ (recall that LBP is iterative) to finally denote the message as $\mu_{m \to n}^{(\ell)}(x_k^{(\ell)}, t)$.

\begin{figure}[!t]
\centering
\includegraphics[width=0.47\textwidth]{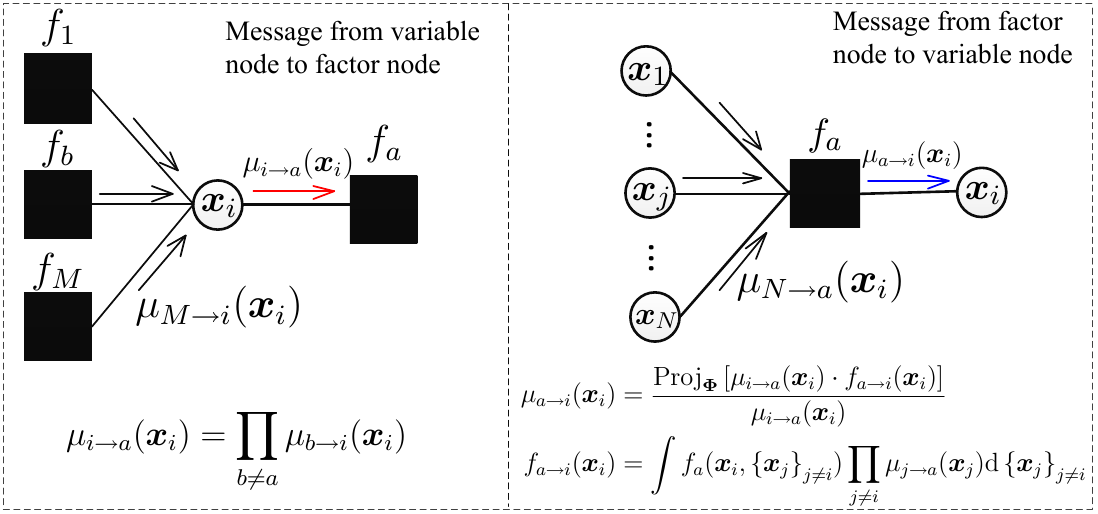}
\caption{Rules for updating messages \cite{zou2018concise}
}
\label{fig:MessageUpdate}
\end{figure}

In these notations, the essential rules for message updating could be depicted as Fig. \ref{fig:MessageUpdate}, following our previous method in  \cite{zou2018concise}. The key idea there was to modify the updating of messages sent from a factor node to a variable node by some moment-matching projection, see the r.h.s. of the figure. The projection was inspired by EP \cite{minka2001family}/EC\cite{opper2005expectation}, and it uses a Kullback-Leibler (KL) divergence  to measure the ``distance'' between two distribution functions. To be specific, let $p(x)$ and $q(x)$ denote two densities, and $\mathcal{D}[q(x) \| p(x)]$ denote the KL divergence from $p(x)$ to $q(x)$. Then, the moment-matching projection given below could offer an approximate function that is Gaussian with a mean and a variance precisely match that of the target function $q(x)$, i.e.,
\begin{align}
\mathrm{Proj}[q({x})]
& \triangleq
\arg \min_{p(x) \in {\Phi}} \mathcal{D} [q({x})\|p({x})]
=
\mathcal{N} ({x}| {m}, {v} )
,
\label{eq:proj_KL}
\end{align}
where ${\Phi}$ denotes all possible Gaussian densities,\\
$m
=
\int {x} q({x}) \, \mathrm{d} {x}
,
$
and
$v
=
    \int
    |{x}-{m}|^2 q({x})
    \, \mathrm{d} {x}
$.

Given the message updating rules, the backbone of the ML-GAMP can now be established, which reads
\begin{subequations}\label{eq:LBP}
\zerospace
\begin{eqnarray}
\mu_{i\leftarrow a}^{(\ell)}(x_i^{(\ell)},t)
&\propto &
    \frac{\text{Proj}_{\bs{\Phi}}[\mu_{i\to a}^{(\ell)}(x_i^{(\ell)},t)  f_{i\leftarrow a}^{(\ell)}(x_i^{(\ell)},t)]}{\mu_{i\to a}^{(\ell)}(x_i^{(\ell)},t)}
\label{A4}\\
\mu_{i\leftarrow i}^{(\ell)}(x_i^{(\ell)},t)
&\propto &
    \prod_{b}\mu_{i\leftarrow b}^{(\ell)}(x_i^{(\ell)},t)
\\
\mu_{i\to i}^{(\ell)}(x_i^{(\ell)},t+1)
&\propto &
    \frac{\text{Proj}_{\bs{\Phi}}[f_{i\to i}^{(\ell)}(x_i^{(\ell)},t+1) \mu_{i\leftarrow i}^{(\ell)}(x_i^{(\ell)},t)]}{\mu_{i\leftarrow i}^{(\ell)}(x_i^{(\ell)},t)}
\label{eq:mu_i2i}
    \quad
\\
\mu_{i\to a}^{(\ell)}(x_i^{(\ell)},t+1)
&\propto &
    \mu_{i\to i}^{(\ell)}(x_i^{(\ell)},t+1) \prod_{b\ne a}\mu_{i\leftarrow b}^{(\ell)}(x_i^{(\ell)},t)
\label{A5}
\end{eqnarray}
\end{subequations}
where
$\propto$ denotes ``is proportional to $x$'',
$\mu_{i\leftarrow a}^{(\ell)}(x_i^{(\ell)},t)$ is the message from the factor node $\mathcal{P}(x_a^{(\ell+1)}|\bs{x}^{(\ell)})$ to the variable node $x_i^{(\ell)}$, and $\mu_{i\to a}^{(\ell)}(x_i^{(\ell)},t)$ is that in the opposite direction.
Similarly, $\mu_{i\leftarrow i}^{(\ell)}(x_i^{(\ell)},t)$ is the message from $x_i^{(\ell)}$ to $\mathcal{P}(x_i^{(\ell)}|\bs{x}^{(\ell-1)})$, while $\mu_{i\to i}^{(\ell)}(x_i^{(\ell)},t+1)$ is the opposite. The following notations are also used in the backbone iteration:
{\zerospace
\begin{align}
f_{i\leftarrow a}^{(\ell)}(x_i^{(\ell)},t)
&\propto
    \int \mathcal{P}(x_a^{(\ell+1)}|\bs{x}^{(\ell)})\mu_{a\leftarrow a}^{(\ell)}(x_a^{(\ell+1)},t)
\nonumber\\ & \quad \quad
    \prod_{j\ne i}\mu_{j\to a}^{(\ell)}(x_j^{(\ell)},t)
\dd \bs{x}_{\backslash i}^{(\ell)}\dd x_a^{(\ell+1)}
\label{eq:f_a2i}
\\
\mu_{a\leftarrow a}^{(\ell)}(x_a^{(\ell+1)},t)
&\propto
    \prod_{k=1}^{N_{\ell+2}}\mu_{a\leftarrow k}^{(\ell+1)}(x_a^{(\ell+1)},t)
\label{eq:mu_a2a}
\\
f_{i\to i}^{(\ell)}(x_i^{(\ell)},t+1)
& \propto
    \!\! \int \!\! \mathcal{P}(x_i^{(\ell)}|\bs{x}^{(\ell-1)})
    \!\!  \prod_{k=1}^{N_{\ell-1}} \!\! \mu_{k\to i}^{(\ell-1)}(x_k^{(\ell)},t+1)\dd \bs{x}^{(\ell-1)}
\end{align}
}
Then, we combine (\ref{eq:LBP}) to get a new compacter set of iteration
\begin{subequations}
\zerospace
\begin{eqnarray}
\mu_{i\leftarrow a}^{(\ell)}(x_i^{(\ell)},t)
&\propto &
    \frac{\text{Proj}_{\bs{\Phi}}[\mu_{i\to a}^{(\ell)}(x_i^{(\ell)},t)f_{i\leftarrow a}^{(\ell)}(x_i^{(\ell)},t)]}{\mu_{i\to a}^{(\ell)}(x_i^{(\ell)},t)}
\label{AA1}\\
\mu_{i\to a}^{(\ell)}(x_i^{(\ell)},t+1)
&\propto &
    \frac{\text{Proj}_{\bs{\Phi}}[f_{i\to i}^{(\ell)}(x_i^{(\ell)},t)\prod_{b}\mu_{i\leftarrow b}^{(\ell)}(x_i^{(\ell)},t)]}{\mu_{i\leftarrow a}^{(\ell)}(x_i^{(\ell)},t)}
    \quad\quad
\label{AA2}
\end{eqnarray}
\end{subequations}
where the approximate marginal of $x_i$ at the $T$-th iteration is:
\begin{eqnarray}
\hat{p}(x_i|\bs{y})
& \propto &
\mu_{i\to a}^{(1)}(x_i, T)\cdot \mu_{i\leftarrow a}^{(1)}(x_i, T)
\end{eqnarray}
As LBP could start up in any part of the network, we begin ours with the last layer, i.e., the rightmost block of the factor graph in Fig. \ref{fig:FactoGraph}. Our equations then iterate like this:
the messages, starting from the last layer, propagate layer by layer in a backward direction ($\leftarrow$) to the very first on the leftmost side; and then, they turn around to propagate in the forward direction ($\to$), also layer by layer. When one round trip is done, we call it an iteration, and an estimate could be obtained, although for better estimates, many rounds are needed usually. For a better corporation with the moment matching, we require all the messages to be initialized as Gaussian, which is distinctly different from standard LBP, where no such requirement is made.

Given (\ref{AA1})-(\ref{AA2}), there are still other challenge facing our iteration, and a most urgent one is the computational complexity. Inherited from LBP, there are $O(N^(\ell+1)\times N^(\ell))$ messages in the system needed to be updated per layer per iteration and in a single direction. In order to reduce this message number, previously AMP \cite{donoho2010message} and GAMP \cite{Rangan-arxiv10-GAMP} relied on the Gaussian approximation and Taylor expansion, while in our previous paper \cite{zou2018concise}, we used the above moment-matching projection and demonstrated that the AMP and GAMP algorithm could also be recovered while Taylor expansions that are unfriendly to complex settings could be avoided. In this context we continue to use the method and derive a new algorithm, starting from (\ref{AA1})-(\ref{AA2}). 
Our derivation is then divided into six parts: the first three are for the evaluation of (\ref{AA1}), and the subsequent two are for (\ref{AA2}), while the last one will put all pieces together to given a complete ML-GAMP as in Algorithm \ref{alg:CML-GAMP}.

\noindent\textbf{\underline{Step 1}}: Evaluate $f_{i\leftarrow a}^{(\ell)}(x_i^{(\ell)},t)$ in (\ref{AA1}).\\
First of all, we recall a fact that all messages were initialized as Gaussian, and thus, after the moment-matching projection, all messages in the network will eventually become Gaussian, which is different from standard LBP where the messages are non-Gaussian in general. Noticing this, we then denote
\begin{align}
\mu_{a\leftarrow k}^{(\ell)}(x_a^{(\ell+1)},t)
& \triangleq
\mathcal{N}(x_a^{(\ell+1)}|m_{a\leftarrow k}^{(\ell+1)}(t),v_{a\leftarrow k}^{(\ell+1)}(t))
,
\end{align}
and apply the Gaussian reproduction property%
\footnote{
Gaussian reproduction property \cite{Rasmussen-Book06-GaussianReproduction}:
$
\mathcal{N}(x|a,A)\mathcal{N}(x|b,B)=\mathcal{N}(0|a-b,A+B)\mathcal{N}(x|c,C)$ with $C=\left(\frac{1}{A}+\frac{1}{B}\right)^{-1}$ and $c=C\left(\frac{a}{A}+\frac{b}{B}\right)$.
} %
to (\ref{eq:mu_a2a}) to get
\begin{align}
\mu_{a\leftarrow a}^{(\ell)}(x_a^{(\ell+1)},t)
&=
    \mathcal{N}(x_a^{(\ell)}|R_a^{(\ell+1)}(t),\Sigma_a^{(\ell)}(t))
\\
\Sigma_a^{(\ell+1)}(t)
&=
    \left(\sum_{k=1}^{N_{\ell+2}}\frac{1}{v_{a\leftarrow k}^{(\ell+1)}(t)}\right)^{-1}
\\
R_a^{(\ell+1)}(t)
&=
    \Sigma_a^{(\ell+1)}(t) \cdot \sum_{k=1}^{N_{\ell+2}}\frac{m_{a\leftarrow k}^{(\ell+1)}(t)}{v_{b\leftarrow k}^{(\ell+1)}(t)}
\end{align}
After that, we start the evaluation of $f_{i\leftarrow a}^{(\ell)}(x_i^{(\ell)},t)$, and an interim result is given in (\ref{AA6}) at the top of next page.
\begin{figure*}[!t]
\normalsize
\begin{align}
&f_{i\leftarrow a}^{(\ell)}(x_i^{(\ell)},t)
\propto
    \int \delta\left(z_a^{(\ell)}-\sum_{j=1}^{N_{\ell}}H_{aj}^{(\ell)}x_j^{(\ell)} \right)
    \left(\int \mathcal{P}(x_a^{(\ell+1)}|z_a^{(\ell)})\mu_{a\leftarrow a}^{(\ell)}(x_a^{(\ell+1)},t)\dd x_a^{(\ell+1)}\right)
    \prod_{j\ne i}\mu_{j\to a}^{(\ell)}(x_j^{(\ell)},t)\dd \bs{x}_{\backslash i}^{(\ell)}\dd z_a^{(\ell)}
\\
&\propto
    \int \delta \left( x_i^{(\ell)}-\frac{1}{H_{ai}^{(\ell)}}(z_a^{(\ell)}-\sum_{j\ne i}H_{aj}^{(\ell)}x_j^{(\ell)}) \right)
    \left(\int \mathcal{P}(x_a^{(\ell+1)}|z_a^{(\ell)})\mu_{a\leftarrow a}^{(\ell)}(x_a^{(\ell+1)},t)\dd x_a^{(\ell+1)}\right)
    \prod_{j\ne i}\mu_{j\to a}^{(\ell)}(x_j^{(\ell)},t)\dd \bs{x}_{\backslash i}^{(\ell)}\dd z_a^{(\ell)}
\label{AA6}
\end{align}
\hrule
\end{figure*}
To proceed further, we also notice that although the original LBP treats the $f_{i\leftarrow a}^{(\ell)}(x_i^{(\ell)},t)$ as a likelihood function in $\bs{y}$ conditioned on $x_i^{(\ell)}$, this function in our context could be treated as an p.d.f. in $x_i^{(\ell)}$ with a difference up to some scaling factor depending on $\bs{y}$. To be specific, the function as seen in (\ref{eq:f_a2i}) will eventually take a form like $\alpha(\bs{y}) \cdot \mathcal{N}(x_i | \beta(\bs{y}), \gamma(\bs{y}))$, thanks to the extraordinary  property of self-conjugation of the Gaussian distribution, see \cite{bishop2006pattern} for more information on exponential family, exponential conjugate, and the Gaussian as a unique case in the former two.
Treating this as an p.d.f. in $x_i^{(\ell)}$ substantially facilitates%
\footnote{It is possible to evaluate (approximate) $f_{i\leftarrow a}^{(\ell)}(x_i^{(\ell)},t)$ without treating it as an p.d.f. in $x_i^{(\ell)}$. For instance, in our previous paper \cite{zou2018concise}, we adopted a method that was similar to that in the Bi-GAMP case \cite{parker2014bilinear}, which also relied on the central limit theorem but did not require it to be a p.d.f..
} %
the analysis that follows, where the evaluation will be shifted to an r.v.s perspective (from the p.d.f.'s), as many well-known results in high dimensional random theories were given in a r.v. point of view. To this end, we let ``$\sim$'' stand for ``follows the distribution of'', and introduce new r.v.s to associate with all messages (distributions) involved in the evaluation, including:
$\xi_{j\to a}^{(\ell)}(t)\sim \mu_{j\to a}^{(t)}(x_j^{(\ell)},t)$,
$\zeta_a^{(\ell)}(t)
\sim
\int \mathcal{P}(x_a^{(\ell+1)}|z_a^{(\ell)})\mu_{a\leftarrow a}^{(\ell)}(x_a^{(\ell+1)},t)\dd x_a^{(\ell+1)}$,
and
$\xi_{i\leftarrow a}^{(\ell)}(t)\sim f_{i\leftarrow a}^{(\ell)}(x_i^{(\ell)},t)$.
By that, we could express the distribution of a compound vector $(\zeta_a^{(\ell)}(t),\{\xi_{j\to a}^{(\ell)}(t)\}_{j\ne i})$ as below
\begin{align*}
&
\mathcal{P}
_{\zeta_a^{(\ell)}(t),\{\xi_{j\to a}^{(\ell)}(t)\}_{j\ne i}}
\left(z_a^{(\ell)}, \{x_j^{(\ell)}\}_{j\ne i}\right)
\nonumber\\
&  \propto
    \int \mathcal{P}(x_a^{(\ell+1)}|z_a^{(\ell)})\mu_{a\leftarrow a}^{(\ell)}(x_a^{(\ell+1)},t)\dd x_a^{(\ell+1)}
    \prod_{j\ne i}\mu_{j\to a}^{(\ell)}(x_j^{(\ell)},t)
\end{align*}
Comparing this to (\ref{eq:f_a2i}), we find, after noticing a generic relation%
\footnote{
Given two r.v.s, $\bs{w}\in \mathbb{C}^{k}$ and $u\in \mathbb{C}$, where $\bs{w}\sim \mathcal{P}_{\bs{w}}(\bs{w})$, and for any mapping $g:\mathbb{C}^k \mapsto \mathbb{C}^1$, it holds:
$u=g(\bs{w})$ if and only if $p_u(u)\propto \int \delta(u-g(\bs{w}))\mathcal{P}_{\bs{w}}(\bs{w})\dd \bs{w}$
(see, e.g., \cite{Pfister-14-CompressedSensing}).
} %
between an r.v. and its p.d.f, that the r.v. $\xi_{i\leftarrow a}^{(\ell)}(t)$ associated with $f_{i\leftarrow a}^{(\ell)}(x_i^{(\ell)},t)$ could be reexpressed as a function of the compound random vector above, i.e.,
\begin{align}
\xi_{i\leftarrow a}^{(\ell)}(t)=\frac{1}{H_{ai}^{(\ell)}}\left(\zeta_a^{(\ell)}(t)-\sum_{j\ne i}H_{aj}^{(\ell)}\xi_{j\to a}^{(\ell)}(t)\right)
.
\label{A6}
\end{align}
As $N_{\ell}\to \infty$, the central limit theorem comes into play, thus
\begin{align}
\sum_{j\ne i}H_{aj}^{(\ell)}\xi_{j\to a}^{(t)}
&\sim
    \mathcal{N}(Z_{i\leftarrow a}^{(\ell)}(t),V_{i\leftarrow a}^{(\ell)}(t))
\label{eq:Gauss}
\\
Z_{i\leftarrow a}^{(\ell)}(t)
&=\sum_{j\ne i}H_{aj}^{(\ell)}m_{j\to a}^{(\ell)}(t)
\label{AA7}\\
V_{i\leftarrow a}^{(\ell)}(t)
&=\sum_{j\ne i}|H_{aj}^{(\ell)}|^2v_{j\to a}^{(\ell)}(t)
\label{AA8}
\end{align}
with $m_{j\to a}^{(\ell)}(t)$ and $v_{j\to a}^{(\ell)}(t)$ denoting the mean and the variance of the Gaussian  message $\mu_{j\to a}^{(\ell)}(x_j^{(\ell)},t)$.
Going back to the r.h.s. of (\ref{A6}), since its 2nd term could be approximate by an Gaussian r.v. distributed as (\ref{eq:Gauss}) and its 1st term was given by
$\zeta_a^{(\ell)}(t) \sim \int \mathcal{P}(x_a^{(\ell+1)}|z_a^{(\ell)})\mu_{a\leftarrow a}^{(\ell)}(x_a^{(\ell+1)},t)\dd x_a^{(\ell+1)}$,
we could simply convolute%
\footnote{Given two independent r.v.s, $X\sim p_X(x)$ and $Y\sim p_Y(y)$, the p.d.f. of a third one, $Z=X+Y$, is then given by $p_Z(z)=\int p_X(x)p_Y(z-x)\dd x$.
} %
the p.d.f.'s of these two and get $\xi_{i\leftarrow a}^{(\ell)}(t)$'s
\begin{align*}
f_{i\leftarrow a}^{(\ell)}(x_i^{(\ell)},t)\propto \mathbb{E}_{\zeta_a^{(\ell)}(t)}\left[
    \mathcal{N}(x_i^{(\ell)}|\frac{\zeta_a^{(\ell)}(t)-Z_{i\leftarrow a}^{(\ell)}(t)}{H_{ai}^{(\ell)}},\frac{V_{i\leftarrow a}^{(\ell)}(t)}{|H_{ai}^{(\ell)}|^2})
\right]
\end{align*}

\noindent\textbf{\underline{Step 2}}: Evaluate $\text{Proj}[\mu_{i\to a}^{(\ell)}(u_i^{(\ell)},t) f_{i\leftarrow a}^{(\ell)}(u_i^{(\ell)},t)] $ in (\ref{AA1}).
\begin{align}
\text{Proj}_{\bs{\Phi}}
&
\left[\mu_{i\to a}^{(\ell)}(x_i^{(\ell)},t)\cdot f_{i\leftarrow a}^{(\ell)}(x_i^{(\ell)},t)\right]
=
\text{Proj}_{\bs{\Phi}} \{
    \mu_{i\to a}^{(\ell)}(x_i^{(\ell)},t)\cdot
\nonumber\\&
    \mathbb{E}_{\zeta_a^{(\ell)}(t)}[
        \mathcal{N}(x_i^{(\ell)} | \frac{\zeta_a^{(\ell)}(t)-Z_{i\leftarrow a}^{(\ell)}(t)}{H_{ai}^{(\ell)}},\frac{V_{i\leftarrow a}^{(\ell)}(t)}{|H_{ai}^{(\ell)}|^2} )
    ]
\}
\end{align}
We first notice that the term $\text{Proj}_{\bs{\Phi}} \{\cdot\}$ is playing the role of a joint pdf of the input element and the observation vector. In the presence of the moment-matching projection, the term inside $\text{Proj}_{\bs{\Phi}} \{\cdot\}$ could be treated as a marginal p.d.f. in $x_i^{(\ell)}$.
In this context, we denote the Gaussian message above as
$\mu_{i\to a}^{(\ell)}(x_i^{(\ell)},t) \triangleq \mathcal{N}(x_i^{(\ell)}|m_{i\to a}^{(\ell)}(t),v_{i\to a}^{(\ell)}(t))$,
and use a Gaussian reproduction property to simplify the marginal as
\begin{align}
&
\mu_{i\to a}^{(\ell)}(x_i^{(\ell)},t)\cdot
\mathbb{E}_{\zeta_a^{(\ell)}(t)}
[\mathcal{N}
    (x_i^{(\ell)} | \frac{\zeta_a^{(\ell)}(t)-Z_{i\leftarrow a}^{(\ell)}(t)}{H_{ai}^{(\ell)}},\frac{V_{i\leftarrow a}^{(\ell)}(t)}{|H_{ai}^{(\ell)}|^2} )
]
\label{eq:muPDF}
\\
&=
\mathbb{E}_{\zeta_a^{(\ell)}(t)}[
    \mathcal{N}(x_i^{(\ell)}|\tilde{m}_{i\leftarrow a}^{(\ell)}(t,\zeta_a^{(\ell)}(t)),\tilde{v}_{i\leftarrow a}^{(\ell)}(t))
    \times
\nonumber\\ & \quad \;
    \mathcal{N}(0 |m_{i\to a}^{(\ell)}(t)-\frac{\zeta_a^{(\ell)}(t)-Z_{i\leftarrow a}^{(\ell)}(t)}{H_{ai}^{(\ell)}},v_{i\leftarrow a}^{(\ell)}(t)+\frac{V_{i\leftarrow a}^{(\ell)}(t)}{|H_{ai}^{(\ell)}|^2} )
]
\nonumber\\
&\propto
\mathbb{E}_{\zeta_a^{(\ell)}}[
    \mathcal{N}(x_i^{(\ell)}|\tilde{m}_{i\leftarrow a}^{(\ell)}(\zeta_a^{(\ell)}),\tilde{v}_{i\leftarrow a}^{(\ell)})
    \mathcal{N}(\zeta_a^{(\ell)} | Z_a^{(\ell)}, V_a^{(\ell)})
]
\label{AA3}
\end{align}
where we drop all the indices $(t)$ from the iterating variables in (\ref{AA3}) as they could be recovered easily from the context, and
\begin{align}
\tilde{v}_{i\leftarrow a}^{(\ell)}
&=
    \left(\frac{1}{v_{i\to a}^{(\ell)}}+\frac{|H_{ai}^{(\ell)}|^2}{V_{i\leftarrow a}^{(\ell)}}\right)^{-1}
\\
\!\!\!\!
\tilde{m}_{i\leftarrow a}^{(\ell)}(\zeta_a^{(\ell)})
&=
    \tilde{v}_{i\leftarrow a}^{(\ell)} \left(\frac{m_{i\to a}^{(\ell)}}{v_{i\to a}^{(\ell)}}
+\frac{(H_{ai}^{(\ell)})^{*}(\zeta_a^{(\ell)}-Z_{i\leftarrow a}^{(\ell)})}{V_{i\leftarrow a}^{(\ell)}}
\right)
\\
Z_a^{(\ell)}
&=
    H_{ai}^{(\ell)}m_{i\leftarrow a}^{(\ell)}+Z_{i\leftarrow a}^{(\ell)}
\label{AB3}
\\
V_{a}^{(\ell)}
&=
    |H_{ai}^{(\ell)}|^2v_{i\to a}^{(\ell)}+V_{i\leftarrow a}^{(\ell)}
\label{AB2}
\end{align}
As just mentioned, Eq. (\ref{eq:muPDF}) could be treated as an p.d.f. in $x_i^{(\ell)}$. Since an p.d.f. integrates to unity, we recover the normalizing factor (a.k.a. the partition function) omitted in (\ref{AA3}) to provide below a more complete form of  (\ref{eq:muPDF})
\begin{align}
&\!\!
\frac{
    \mathbb{E}_{\zeta_a^{(\ell)}}[
    \mathcal{N}(\zeta_a^{(\ell)}|Z_a^{(\ell)},V_a^{(\ell)}) \mathcal{N}(x_i^{(\ell)}|\tilde{m}_{i\leftarrow a}^{(\ell)}(t,\zeta_a^{(\ell)}),\tilde{v}_{i\leftarrow a}^{(\ell)})
    ]
}{
    \mathbb{E}_{\zeta_a^{(\ell)}}[
    \mathcal{N}(\zeta_a^{(\ell)}|Z_a^{(\ell)},V_a^{(\ell)})
    ]
}
\label{AA5}
\end{align}
For (\ref{AA5}), we still need to know its mean and variance before evaluating the moment-matching projection $\text{Proj}_{\bs{\Phi}} \{\cdot\}$. This starts by introducing yet another r.v.-to-p.d.f. association
\begin{align}
\tilde{\zeta}_a^{(\ell)}(t)
\sim
\mathcal{P}_{\tilde{\zeta}_a^{(\ell)}}(z_a^{(\ell)})
&\triangleq
\frac{\int \mathcal{N}_{x_a^{(\ell+1)}|z_a^{(\ell)}}(\cdot) \dd x_a^{(\ell+1)}}{\int \mathcal{N}_{x_a^{(\ell+1)}|z_a^{(\ell)}}(\cdot)\dd x_a^{(\ell+1)}\dd z_a^{(\ell)}}
\label{AA4}
\\
\mathcal{N}_{x|z}(a,A,b,B)
&\triangleq
\mathcal{P}(x|z)\mathcal{N}(x|a,A)\mathcal{N}(z|b,B)
\end{align}
where
$(\cdot)=
(R_a^{(\ell+1)}(t),\Sigma_a^{(\ell+1)}(t),Z_a^{(\ell)}(t),V_a^{(\ell)}(t))
$, and the mean and variance of $\tilde{\zeta}_a^{(\ell)}(t)$ are denoted as $\tilde{z}_a^{(\ell)}(t)$ and $\tilde{v}_a^{(\ell)}(t)$:
\begin{align}
\tilde{z}_a^{(\ell)}(t)&=\mathbb{E}_{\tilde{\zeta}_a^{(\ell)}(t)} [\tilde{\zeta}_a^{(\ell)}(t) ]\\
\quad \tilde{v}_a^{(\ell)}(t)&=\text{Var}_{\tilde{\zeta}_a^{(\ell)}(t)} [\tilde{\zeta}_a^{(\ell)}(t) ]
\end{align}
Particularly, if $\ell=L$, the r.h.s. of (\ref{AA4}) becomes
$\frac{\mathcal{P}(y_a|z_a^{(L)})\mathcal{N}(z_a^{(L)}|Z_a^{(L)}(t),V_a^{(L)}(t))}
{\int \mathcal{P}(y_a|z_a^{(L)})\mathcal{N}(z_a^{(L)}|Z_a^{(L)}(t),V_a^{(L)}(t))\dd z_a^{(L)}}$.
In these notations, we are able to express the mean and the variance of (\ref{AA5}) as
\begin{align*}
&
m_i^{(\ell)}
\!\!=\!\!
\int \!\! x_i^{(\ell)} \!\!\!\! \int \!\! \mathcal{P}_{\tilde{\zeta}_a^{(\ell)}}(z_a^{(\ell)})\mathcal{N}(x_i^{(\ell)}|\tilde{m}_{i\leftarrow a}^{(\ell)}(t,z_a^{(\ell)}),\tilde{v}_{i\leftarrow a}^{(\ell)})\dd z_a^{(\ell)}\dd x_i^{(\ell)}
\\
&
=\left(\frac{1}{v_{i\leftarrow a}^{(\ell)}}+\frac{|H_{ai}^{(\ell)}|^2}{V_{i\leftarrow a}^{(\ell)}}\right)^{-1}
\left(\frac{m_{i\to a}^{(\ell)}}{v_{i\to a}^{(\ell)}}+\frac{(H_{ai}^{(\ell)})^{*}(\tilde{z}_a^{(\ell)}-Z_{i\leftarrow a}^{(\ell)})}{V_{i\leftarrow a}^{(\ell)}}\right)
\\
&v_i^{(\ell)}
=\frac{v_{i\to a}^{(\ell)}V_{i\leftarrow a}^{(\ell)}V_a^{(\ell)}+(v_{i\to a}^{(\ell)})^2|H_{ai}^{(\ell)}|^2\tilde{v}_a^{(\ell)}}{(V_a^{(\ell)})^2}
\end{align*}
where all iterating variables have the same iteration index $(t)$. So far, we could express the projection as
\begin{align*}
\text{Proj}_{\bs{\Phi}}[\mu_{i\to a}^{(\ell)}(x_i^{(\ell)},t)\cdot f_{i\leftarrow a}^{(\ell)}(x_i^{(\ell)},t)]
&=
\mathcal{N}(x_i^{(\ell)}|m_i^{(\ell)}(t),v_i^{(\ell)}(t))
\end{align*}

\noindent\textbf{\underline{Step 3}}: Evaluate $\mu_{i\leftarrow a}^{(\ell)}(x_i^{(\ell)},t+1)$ in (\ref{AA1}). \\
By Gaussian reproduction property, the mean and variance of $\mu_{i\leftarrow a}^{(\ell)}(x_i^{(\ell)},t+1)$ could be expressed as
\begin{align}
&v_{i\leftarrow a}^{(\ell)}(t+1)=
    %
    %
    \frac{(V_a^{(\ell)})^2-v_{i\to a}^{(\ell)}|H_{ai}^{(\ell)}|^2(V_a^{(\ell)}-\tilde{v}_a^{(\ell)})}{|H_{ai}^{(\ell)}|^2(V_a^{(\ell)}-\tilde{v}_a^{(\ell)})}
\\
&m_{i\leftarrow a}^{(\ell)}(t+1)=
\nonumber\\&
    %
    %
    \frac{(H_{ai}^{(\ell)})^{*}(\tilde{z}_a^{(\ell)}-Z_a^{(\ell)})V_a^{(\ell)}+|H_{ai}^{(\ell)}|^2m_{i\to a}^{(\ell)}(V_a^{(\ell)}-\tilde{v}_a^{(\ell)})}
    {|H_{ai}^{(\ell)}|^2(V_a^{(\ell)}-\tilde{v}_a^{(\ell)})}
\end{align}
where the index $(t)$ is omitted from all iterating variables. Further ignoring the high-order items, we could get
\begin{align}
v_{i\leftarrow a}^{(\ell)}(t)&=\frac{1}{|H_{ai}|^2\tau_a^{(\ell)}(t)}\\
m_{i\leftarrow a}^{(\ell)}(t)&=\frac{(H_{ai}^{(\ell)})^{*}s_a^{(\ell)}(t)+|H_{ai}^{(\ell)}|^2\tau_a^{(\ell)}(t)m_{i\to a}^{(\ell)}(t)}{|H_{ai}^{(\ell)}|^2\tau_a^{(\ell)}(t)}
\end{align}
where the followings are defined
\begin{align}
\tau_a^{(\ell)}(t)&\triangleq \frac{V_a^{(\ell)}(t)-\tilde{v}_a^{(\ell)}(t)}{(V_a^{(\ell)}(t))^2}\\
s_a^{(\ell)}(t)&\triangleq \frac{\tilde{z}_a^{(\ell)}(t)-Z_a^{(\ell)}(t)}{V_a^{(\ell)}(t)}
\end{align}

\noindent\textbf{\underline{Step 4}}: Do $\text{Proj}[f_{i\to i}^{(\ell)}(x_i^{(\ell)} \!\!,t+1)\prod_{b}\mu_{i\leftarrow b}^{(\ell)}(x_i^{(\ell)} \!\!,t)]$ in (\ref{AA2}). \\
We first evaluate $f_{i\to i}^{(\ell)}(x_i^{(\ell)},t+1)$ in a way similar to (\ref{AA6})-(\ref{AA8}),
\begin{align}
f_{i\to i}^{(\ell)}
&
(x_i^{(\ell)},t+1)
\propto
    \int \dd z_i^{(\ell-1)} \mathcal{P}(x_i^{(\ell)}|z_i^{(\ell-1)}) \times
\nonumber\\
&
    \mathcal{N}(z_i^{(\ell-1)}|Z_i^{(\ell-1)}(t+1),V_i^{(\ell-1)}(t+1))
\end{align}
By Gaussian reproduction property, we further get
\begin{align*}
\text{Proj}_{\bs{\Phi}}\left[f_{i\to i}^{(\ell)}\prod_{b}\mu_{i\leftarrow b}^{(\ell)}\right]
&\propto
    \text{Proj}_{\bs{\Phi}}\left[
    \int \mathcal{N}_{x_i^{(\ell)}|z_i^{(\ell-1)}}(\cdot) \dd z_i^{(\ell-1)}
    \right]
\\
&\propto
    \mathcal{N}(x_i^{(\ell)}|\hat{m}_i^{(\ell)}(t+1),\hat{v}_i^{(\ell)}(t+1))
\end{align*}
where the following definitions are used
\begin{align}
& \!\!\!\!
(\cdot)
=
    (
    R_i^{(\ell)}(t),\Sigma_i^{(\ell)}(t),Z_i^{(\ell-1)}(t+1),V_i^{(\ell-1)}(t+1)
    )
\\
& \!\!\!\!
\Sigma_i^{(\ell)}(t)
\triangleq
    [
    \sum\nolimits_{a=1}^{N_{\ell+1}}|H_{ai}^{(\ell)}|^2\tau_a^{(\ell)}(t)
    ]^{-1}
\label{AB5}
\\
& \!\!\!\!
R_i^{(\ell)}(t)
\!\triangleq\!\!
    \Sigma_i^{(\ell)}(t)
    \!\!\!\sum_{a=1}^{N_{\ell+1}}\!\! (\!H_{ai}^{(\ell)})^{*}s_a^{(\ell)}(t)
    \!\!+\!\!
    |H_{ai}^{(\ell)}|^2\tau_a^{(\ell)}(t)m_{i\to a}^{(\ell)}(t)
\label{AB6}
\\
& \!\!\!\!
\hat{m}_i^{(\ell)}(t+1)
\triangleq
    \mathbb{E}[ \xi_i^{(\ell)}(t+1)]
\\
& \!\!\!\!
\hat{v}_i^{(\ell)}(t+1)
\triangleq
    \text{Var}[\xi_i^{(\ell)}(t+1)]
\end{align}
with the expectation and variance in the last two equalities being taken from this p.d.f. of $\xi_i^{(\ell)}(t+1)$
\begin{align}
\mathcal{P}_{\xi_i^{(\ell)}(t+1)}(x_i^{(\ell)})
& =
    \frac{
        \int \mathcal{N}_{x_i^{(\ell)}|z_i^{(\ell-1)}}(\cdot) \dd z_i^{(\ell-1)}
    }{
        \int \mathcal{N}_{x_i^{(\ell)}|z_i^{(\ell-1)}}(\cdot) \dd z_i^{(\ell-1)}\dd x_i^{(\ell)}
    }
\end{align}
In the particular case of $\ell=1$, we initialize for consistency $x_i^{(1)}:=x_i$
and
$
\mathcal{P}_{\xi_i^{(1)}(t+1)}(x_i^{(1)})
:=
\frac{\mathcal{P}_X(x_i)\mathcal{N}(x_i|R_i^{(1)},\Sigma_i^{(1)})}{\int \mathcal{P}_X(x)\mathcal{N}(x|R_i^{(1)},\Sigma_i^{(1)}) \dd x}$.

\noindent\textbf{\underline{Step 5}}: Evaluete $\mu_{i\to a}^{(\ell)}(x_i^{(\ell)},t+1)$ in (\ref{AA2}).\\
Applying the Gaussian reproduction property and then ignoring the high-order infinitesimals, we get
\begin{align}
v_{i\to a}^{(\ell)}(t+1)
&=
    \left(
    \frac{1}{\hat{v}_i^{(\ell)}(t+1)}-\frac{1}{v_{i\leftarrow a}^{(\ell)}(t+1)}
    \right)^{-1}
\\
&=
    \hat{v}_i^{(\ell)}(t+1)
\label{AB1}\\
m_{i\to a}^{(\ell)}(t+1)
&=v_{i\to a}^{(\ell)}(t+1)\left(\frac{\hat{m}_i^{(\ell)}(t+1)}{\hat{v}_i^{(\ell)}(t+1)}-\frac{m_{i\leftarrow a}^{(\ell)}(t+1)}{v_{i\leftarrow a}^{(\ell)}(t+1)}\right)
\nonumber\\
& =
\hat{m}_i^{(\ell)}(t+1)
\! - \!
(H_{ai}^{(\ell)})^*s_a^{(\ell)}(t)\hat{v}_i^{(\ell)}(t+1)
\label{AB4}
\end{align}

\noindent\textbf{\underline{Step 6}}: Evaluate $(Z_a^{(\ell)}(t),V_a^{(\ell)}(t),R_i^{(\ell)}(t))$ and gather all.\\
We substitute (\ref{AB1})-(\ref{AB4}) into (\ref{AB3})-(\ref{AB2}) to get
\begin{align*}
V_a^{(\ell)}(t+1)
&=
    \sum_{i=1}^N |H_{ai}^{(\ell)}|^2\hat{v}_i^{(\ell)}(t+1)
\\
Z_a^{(\ell)}(t+1)
&=
    \sum_{i=1}^N H_{ai}^{(\ell)}\hat{m}_i^{(\ell)}(t+1)-s_a^{(\ell)}(t)V_a^{(\ell)}(t+1)
\end{align*}
Substituting (\ref{AB5}) and (\ref{AB4}) into (\ref{AB6}) further yields
\begin{align}
R_i^{(\ell)}(t)=\hat{m}_i^{(\ell)}(t)+\Sigma_i^{(\ell)}(t)\sum_{a=1}^M(H_{ai}^{(\ell)})^{*}s_a^{(\ell)}(t)
\end{align}
So far, we have finished the evaluation of the two messages given by (\ref{AA1})-(\ref{AA2}). From their results we see that these messages are fully characterized by a set of eight iterating variables, i.e., ($\tilde{z}_a^{(\ell)}(t),\tilde{v}_a^{(\ell)}(t)$,$R_i^{(\ell)}(t),\Sigma_i^{(\ell)}(t)$,
$\hat{m}_i^{(\ell)}(t+1),\hat{v}_i^{(\ell)}(t+1)$,$V_a^{(\ell)}(t+1),Z_a^{(\ell)}(t+1)) $, each depending on either $a$ or $i$. Comparing to each original message in (\ref{AA1})-(\ref{AA2}) which depends on $a$ and $i$, the new iteration could effectively reduce the number of messages per layer from $\mathcal{O}(N_{\ell+1}\times N_{\ell})$ to $\mathcal{O}(N_{\ell+1}+N_{\ell})$. So, we re-arrange them in a backward-first manner, and finally recover ML-GAMP as Algorithm.~\ref{alg:CML-GAMP}.

\subsection{Convergence of the Algorithm}
The original AMP \cite{donoho2009message} since its proposal by Donoho \emph{et al.} had been observed to exhibit a fast convergence in the case of large i.i.d. zero-mean Gaussian weighting matrices. However, its convergence under general transform matrices was not fully understood for a long time until recently Rangan \emph{et al.} \cite{Rangan-TIT19-Convergence_GAMP} provided some sufficient conditions, considering damped GAMP (i.e., an extension to AMP) with quadratic cost functions. It was shown there, with sufficient damping, the GAMP algorithm was guaranteed to converge, although the amount of damping grows with peak-to-average ratio of the squared singular values of the transform matrices. Their result explains not only the good performance of the original AMP on i.i.d. Gaussian transform matrices, but also the difficulties in applying the original AMP to ill-conditioned or non-zero-mean transform matrices.

Inherited from AMP and GAMP, our proposed ML-GAMP also exhibits some fast convergence in large i.i.d. zero-mean Gaussian matrices, as we will see later in the experiment section. This is one important aspect in common, but differences also exit. For example, we do find that as the layer number increases, the need for a damping quickly ramps up. Due to  time and space limitation, here we are not able to provide a thorough investigation into this problem, but below we summarize some rules observed in our earlier experience with the algorithm:\\
$\bullet$
Among them, these two variables could be damped first, i.e., $Z_a^{(\ell)}$ in (\ref{AD9}) and $V_a^{(\ell)}$ in (\ref{AD10}), and their damping reads
\begin{align*}
Z_a^{(\ell)}(t+1)
=&
    \rho^{(\ell)}[\sum\nolimits_iH_{ai}^{(\ell)}\hat{m}_i^{(\ell)}(t+1)-V_a^{(\ell)}(t+1)s_a^{(\ell)}(t)]
\nonumber\\&
    +
    [1-\rho^{(\ell)}] Z_a^{(\ell)}(t)
\\
V_a^{(\ell)}(t+1)
=&
    \rho^{(\ell)} [\sum\nolimits_i |H_{ai}^{(\ell)}|^2\hat{v}_i^{(\ell)}(t+1)]
    +
    [1-\rho^{(\ell)}] V_a^{(\ell)}(t)
\end{align*}
$\bullet$
The variable $\rho^{(\ell)}$ is a damping factor and could be chosen freely from $(0,1]$ with $\rho^{(\ell)}=1$ meaning no damping at all. From our early experience with the algorithm, we find it better to limit the factor within $[0.7, 1]$.
\\
$\bullet$
If more damping is desired, two more variables could be considered, i.e., $\hat{m}_i^{(\ell)}(t+1)$ and $\tilde{z}_a^{(\ell)}(t)$, but be aware that excessive damping  certainly slows down the convergence rate.

\subsection{Simplification to the Algorithm}
To further ease ML-GAMP's burden in computation, we provide in Algorithm \ref{alg:SV-MLGAMP} a simplified version, scalar-variance ML-GAMP, where the original variance vectors ($\{V_a^{(\ell)}\}_{a=1,2,\cdots}$ and $\{\Sigma_i^{(\ell)}\}_{i=1,2,\cdots}$) are averaged to be scalars (i.e., $V^{(\ell)}$ and $\Sigma^{(\ell)}$).
The averaging is indeed inspired and supported by a \emph{self-averaging property} that had long been observed in high dimensional studies and was also applied to the design of S-AMP \cite{cakmak2014s} and simplified GAMP \cite{Rangan-arxiv10-GAMP}.

To derive the scalar-variance ML-GAMP, we first replace the two variances $\hat{v}_i^{(\ell)}(t)$ and $\tilde{v}_a^{(\ell)}(t)$ with their own averages:
\begin{align}
\hat{v}_i^{(\ell)}(t)
&\leftarrow
\overline{\hat{v}^{(\ell)}}(t)\triangleq
\frac{1}{N_{\ell}}\sum_{i=1}^{N_{\ell}}\hat{v}_i^{(\ell)}(t)
\\
\tilde{v}_a^{(\ell)}(t)
&\leftarrow \overline{\tilde{v}^{(\ell)}}(t)\triangleq \frac{1}{N_{\ell+1}}\sum_{a=1}^{N_{\ell+1}}\tilde{v}_a^{(\ell)}(t)
\\
V_a^{(\ell)}(t+1)
&\leftarrow
V^{(\ell)}(t+1)
\triangleq
    \frac{1}{\alpha_{\ell}}\overline{\hat{v}^{(\ell)}}(t+1)
,
\label{AE1}
\end{align}
where (\ref{AE1}) is due to
$
V_a^{(\ell)}
\approx |H_{ai}^{(\ell)}|^2\sum\nolimits_i \hat{v}_i^{(\ell)}
\approx \frac{1}{N_{\ell+1}}\sum\nolimits_i \hat{v}_i^{(\ell)}
$.
Substituting the above back into (\ref{AD3}) and (\ref{AD5}) further yields
\begin{align}
s_a^{(\ell)}(t)
& \leftarrow
    \frac{\tilde{z}_a^{(\ell)}(t)-Z_a^{(\ell)}(t)}{V^{(\ell)}(t)}
\\
\Sigma_i^{(\ell)}(t)
& \leftarrow
    \Sigma^{(\ell)}(t)
    \triangleq
    \frac{(V^{(\ell)}(t))^2}{V^{(\ell)}(t)-\overline{\tilde{v}^{(\ell)}}(t)}
\label{III_A1}
\end{align}
After that, we collect all results and come up with
\begin{align*}
R_i^{(\ell)}(t)&=\hat{m}_i^{(\ell)}(t)+\Sigma_i^{(\ell)}(t)\sum_a(H_{ai}^{(\ell)})^{*}s_a^{(\ell)}(t)\\
Z_a^{(\ell)}(t+1)&=\sum_i H_{ai}^{(\ell)}\hat{m}_i^{(\ell)}(t+1)-V_a^{(\ell)}(t+1)s_a^{(\ell)}(t)
\end{align*}
which, after a rearranging in matrix form, recovers Algo. \ref{alg:SV-MLGAMP}.

\begin{algorithm}[!t]
\caption{A Simplified ML-GAMP (Scalar-Variance)}
\label{alg:SV-MLGAMP}
%
\For{$t=1,\cdots,T$ }
{
   \For{$\ell=L,\cdots,1$ }
   {
      \begin{align*}
      \tilde{z}_a^{(\ell)}(t)
      &=\mathbb{E}_{\tilde{\zeta}_a^{(\ell)}(t)}\left[\tilde{\zeta}_a^{(\ell)}(t)\right]
      \\
      \overline{\tilde{v}^{(\ell)}}(t)
      &=\frac{1}{N_{\ell+1}}\sum_{a=1}^{N_{\ell+1}}\text{Var}_{\tilde{\zeta}_a^{(\ell)}(t)}\left[\tilde{\zeta}_a^{(\ell)}(t)\right]
      \\
      s_a^{(\ell)}(t)&=\frac{(\tilde{z}_a^{(\ell)}(t)-Z_a^{(\ell)}(t))}{V^{(\ell)}(t)}
      \\
      \Sigma^{(\ell)}(t)&=\frac{(V^{(\ell)}(t))^2}{V^{(\ell)}(t)-\overline{\tilde{v}^{(\ell)}}(t)}
      \\
      R_i^{(\ell)}(t)&=\hat{m}_i^{(\ell)}(t)+\Sigma^{(\ell)}(t)\sum_a(H_{ai}^{(\ell)})^{*}s_a^{(\ell)}(t)
      \end{align*}
   }
   \For{$\ell=1,\cdots,L$}
   {
    \begin{align*}
    \hat{m}_i^{(\ell)}(t+1)&=\mathbb{E}_{\xi_i^{(\ell)}(t+1)}\left[\xi_{i}^{(\ell)}(t+1)\right]
    \\
    \overline{\hat{v}^{(\ell)}}(t+1)&=\frac{1}{N_{\ell}}\sum_{i=1}^{N_{\ell}}\text{Var}_{\xi_i^{(\ell)}(t+1)}\left[\xi_{i}^{(\ell)}(t+1)\right]
    \\
    V^{(\ell)}(t+1)&=\frac{1}{\alpha_{\ell}}\overline{\hat{v}^{(\ell)}}(t+1)
    \\
    Z_a^{(\ell)}(t+1)&=\sum_i H_{ai}^{(\ell)}\hat{m}_i^{(\ell)}(t+1)-V_a^{(\ell)}(t+1)s_a^{(\ell)}(t)
    \end{align*}
   }
}
\end{algorithm}

\section{ML-GAMP: Asymptotic Analysis and SE} \label{sec:SE}
Inherited from AMP \cite{donoho2009message}, variants in this family usually exhibit a good property that their asymptotic behaviors could be accurately characterized (predicted) through a set of simple one-dimensional iterating equations called \emph{state evolution} (SE).
In the literature, Bayati \emph{et al.} \cite{bayati2011dynamics} rigorously proved the SE of the original AMP, and later Rangan
\cite{Rangan-arxiv10-GAMP} offered its GAMP extension. More recently, Manoel \emph{et al.} presented the SE for their ML-AMP algorithm \cite{manoel2017multi}. Since an SE relies heavily on the implementation of an algorithm (it would capture the per iteration performance of the algorithm), we particularly derive the SE for our ML-GAMP.
The derivation, similar to those given in \cite{Rangan-arxiv10-GAMP} and \cite{manoel2017multi}, are heuristic by nature, while a proof rigorous in the mathematical sense is still pending. But fortunately, the SE derived could capture precisely the per iteration MSE behavior of the ML-GAMP algorithm, as one will see in Fig. \ref{sim_iteration}.
This section is then divided into two parts: the first part presents the SE result, while the second part elaborates all derivation details.


\subsection{Results of the Analysis}

The SE of the ML-GAMP estimator is given in Algorithm \ref{alg:ML-GAMP_SE} under the assumption of a Gaussian weighting matrix $\bs{H}^{(\ell)}$ for all $\ell$, whose elements are i.i.d., zero-mean, and $\mathcal{O}(\frac{1}{N_{\ell+1}})$-variance (to ensure column unity).


\begin{algorithm}[!t]
\caption{State Evolution of the ML-GAMP Estimator
}
\label{alg:ML-GAMP_SE}
\textbf{1. Define:}
\setlength{\arraycolsep}{0pt}
\begin{eqnarray*}
\mathcal{P}^{(\ell)}(x|z) \triangleq \mathcal{P}_{x^{(\ell)}|z^{(\ell-1)}}(x|z)
,
\quad
\mathcal{N}^{(\ell)}_{x|z}(\cdot) \triangleq \mathcal{N}_{x^{(\ell)}|z^{(\ell-1)}}(\cdot)
\end{eqnarray*}
\\
\textbf{2. Initialize:}\\
\For{$\ell=1,\cdots,L$}
{
\begin{eqnarray*}
T_X^{(\ell)}
& = &
    \begin{cases}
    \ell=1:
    &
        \sigma_X^2
    \\
    \ell>1:
    &
        \int |x|^2\mathcal{P}^{(\ell)}(x|z)\mathcal{N}(z|0,T_Z^{(\ell-1)})\dd z\dd x
    \end{cases}
\\
T_Z^{(\ell)}
& = &
    {T_X^{(\ell)}}/{\alpha_{\ell}}
\end{eqnarray*}
}
\textbf{3. Iterate:}\\
\For{$\ell=L,\cdots,1$}
{
\begin{eqnarray*}
&&
V^{(\ell)}
=
    \frac{T_X^{(\ell)}-d^{(\ell)}}{\alpha_{\ell}}
\\
&&q^{(\ell)} =
\nonumber\\
&&
    \begin{cases}
    \ell=L:
    \\
        \int \frac{|\int z \mathcal{P}^{(L)}(y|z)\mathcal{N} (z|\sqrt{\frac{d^{(L)}}{\alpha_{L}}} \xi,\frac{T_X^{(L)}-d^{(L)}}{\alpha_{L}}) \dd z|^2}{\int \mathcal{P}^{(L)}(y|z)\mathcal{N}(z|\sqrt{\frac{d^{(L)}}{\alpha_{L}}} \xi,\frac{T_X^{(L)}-d^{(L)}}{\alpha_{L}} )\dd z}\text{D}\xi\dd y
    \\
    \ell< L:
    \\
        \int\frac{|\int z\mathcal{N}^{(\ell)}_{x|z} (\zeta,\Sigma^{(\ell+1)},\sqrt{T_Z^{(\ell+1)}-V^{(\ell+1)}} \xi,V^{(\ell+1)})
        \dd z\dd x|^2}{\int \mathcal{N}^{(\ell)}_{x|z}(\zeta,\Sigma^{(\ell+1)},\sqrt{T_Z^{(\ell+1)}-V^{(\ell+1)}} \xi,V^{(\ell+1)})
        \dd z\dd x}\text{D}\xi\dd \zeta
    \end{cases}
\\
&&
\Sigma^{(\ell)}
=
    \frac{(V^{(\ell)})^2}{ V^{(\ell)}-T_Z^{(\ell)}+q^{(\ell)} }
\end{eqnarray*}
}

\For{$\ell=1,\cdots,L$}
{
\begin{eqnarray*}
&&d^{(\ell)} =
\nonumber\\
&&
    \begin{cases}
    \ell=1:
    \\ \quad
        \int \frac{|\int x \mathcal{P}_X(x)\mathcal{N}(x|\zeta,\Sigma^{(1)})\dd x|^2}{\int \mathcal{P}_X(x)\mathcal{N}(x|\zeta,\Sigma^{(1)})\dd x}\dd \zeta
    \\
    \ell>1:
    \\
        \int \frac{|\int x \mathcal{N}^{(\ell)}_{x|z}(\zeta,\Sigma^{(\ell-1)},\sqrt{T_Z^{(\ell-1)}-V^{(\ell-1)}} \xi,V^{(\ell-1)})
        \dd z\dd x|^2}
        {\int \mathcal{N}^{(\ell)}_{x|z} (\zeta,\Sigma^{(\ell-1)},\sqrt{T_Z^{(\ell-1)}-V^{(\ell-1)}} \xi,V^{(\ell-1)})
        \dd z\dd x}\text{D}\xi\dd \zeta
    \end{cases}
\end{eqnarray*}

}
\end{algorithm}


For this SE in Algorithm \ref{alg:ML-GAMP_SE}, we have the following remarks:\\
\noindent{\underline{Remark 1}}:
The this SE of ML-GAMP degenerates to many results reported earlier in the literature. For example, at $L=1$, it reduces to the SE of GAMP as given in \cite{rangan2010estimation}:
\begin{align}
\overline{v}&=\frac{\sigma_X^2}{\alpha}-\int \frac{\left|\int zp(y|z)\mathcal{N}(z|\sqrt{\frac{\sigma_X^2-\varepsilon(\Sigma)}{\alpha}}\xi,\frac{\varepsilon(\Sigma)}{\alpha})\dd z\right|^2}{\int \mathcal{P}(y|z)\mathcal{N}(z|\sqrt{\frac{\sigma_X^2-\varepsilon(\Sigma)}{\alpha}}\xi,\frac{\varepsilon(\Sigma)}{\alpha})\dd z}\text{D}\xi\dd y
\nonumber\\
\Sigma&=\frac{(\varepsilon(\Sigma))^2}{\alpha \varepsilon(\Sigma)-\alpha^2\overline{v}}
\label{CA2}
\end{align}
where $\varepsilon(\Sigma)$ is the MSE of a SISO MMSE estimation problem: $Y=X+W$, with $W\sim \mathcal{N}(0,\Sigma)$.
Also at $L=1$, but consider a particular case of SLM, i.e., $\mathcal{P}(y|z)=\mathcal{N}(y|z,\sigma_w^2)$ with $\sigma_w^2$ being a noise power, the above SE further degenerates to the following SE of AMP that was rigorously proved (via a different conditioning technique) in \cite{bayati2011dynamics} :
\begin{align}
\Sigma=\sigma_w^2+\varepsilon(\Sigma)/\alpha
.
\label{CA3}
\end{align}
We also note that the fixed points of our SE agrees perfectly with those of ML-AMP's SE \cite{manoel2017multi}, and this has a multi-folding meaning.
Firstly, a fixed point means that the algorithm has converged and that its averaged MSE has hit the floor. Since the two SE's are sharing the same fixed points, their error floor must be the same, which is indeed confirmed later by Fig. \ref{sim_iteration}.
Secondly, while fixed points define their commonality in the error floor, the differences in (the arrangement of iterating variables of ) the two SE's would sometimes lead to a dramatic change in the dynamical behaviors. This is also evidenced by our later Fig. \ref{sim_iteration}, where, before converged, i.e., before hitting their common error floor, the two algorithms behave dramatically different.
\\
\noindent{\underline{Remark 3}}:
Comparing the above result to Algorithm I of Part I, we find a perfect agreement in fixed points between the SE of ML-GAMP and the replica equations of an exact MMSE estimator.
To see this agreement, we rewrite $V^{(\ell)}$ and $\Sigma^{(\ell)}$ as
\begin{align}
V^{(\ell)}
= \frac{T_X^{(\ell)}-d^{(\ell)}}{\alpha_{\ell}}
\overset{(a)}{\longleftrightarrow}
T_Z^{(\ell)}-V^{(\ell)}=\frac{d^{(\ell)}}{\alpha_{\ell}}
,
\label{eq:relate1}
\\
\Sigma^{(\ell)}
=\frac{(T_X^{(\ell)}-d^{(\ell)})^2}{\alpha_{\ell}(\alpha_{\ell}q^{(\ell)}-d^{(\ell)})}
\overset{(b)}{\longleftrightarrow}
\Sigma^{(\ell)}
=
\frac{1}{2\tilde{d}^{(\ell)}}
,
\label{eq:relate2}
\end{align}
where (a) is due to an earlier equality in the algorithm, i.e.,
$T_Z^{(\ell)}={T_X^{(\ell)}}/{\alpha_{\ell}}$,
while (b) holds as a result of the following equality%
\footnote{
Again, we remind the readers about a difference between iterating equations (e.g., Algorithm \ref{alg:ML-GAMP_SE} here) and coupled equations (e.g., Algorithm 1 in Part I). The SE in Algorithm \ref{alg:ML-GAMP_SE} is a set of iterating equations where the output of earlier steps form the input of what follows. In contrast, the fixed point equation from Part I's Algorithm 1 is just a set of coupled equations (waiting to be solved), meaning that they may not iterate. As a matter of fact, the equality there
$
\tilde{d}^{(\ell)}
=
    \frac{\alpha_{\ell}(\alpha_{\ell} q^{(\ell)}-d^{(\ell)})}{2(T_X^{(\ell)}-d^{(\ell)})^2}
,
$
had opened the iterating chain, because $d^{(\ell)}$ could not be obtained from previous steps unless initialization.
} %
from Part I's Algorithm 1:
$
\tilde{d}^{(\ell)}
=
    \frac{\alpha_{\ell}(\alpha_{\ell} q^{(\ell)}-d^{(\ell)})}{2(T_X^{(\ell)}-d^{(\ell)})^2}
.
$
Incorporating (\ref{eq:relate1}) and (\ref{eq:relate2}) into the SE given in Algorithm \ref{alg:ML-GAMP_SE}, one recovers exactly the fixed point equations of the replica analysis on an exact MMSE estimator, given as Algorithm 1 in Part I.
Since an exact MMSE estimator is Bayes-optimal, such an agreement indicates that the proposed ML-GAMP (if converged) is asymptotically optimal in the MSE sense, and that unlike the exact MMSE estimator, its computational complexity is similar to the GAMP \cite{rangan2011generalized}, and thus affordable.
\\
\noindent{\underline{Remark 4}}:
Since the SE could capture the MSE behavior of an algorithm, one could obtain ML-GAMP's MSE of by simply evaluating its SE. In other words, there is no further need to carry out the time-consuming Monte Carlo simulations which used to model the entire process of data generation, signal transmission, and data estimation.
Moreover, given an average MSE from the SE, the symbol error rate (SER) and many other performance indices could be obtained via closed form transformations. For example, in the case of a  QPSK-distributed $x$, the SER could be expressed analytically as \cite[p. 269]{Proakis-book01-DigitalCommu}:
$
\text{SER}=2Q(\sqrt{\text{MSE}})-[Q(\sqrt{\text{MSE}})]^2
,
$
where $Q(x)=\int_{x}^{+\infty}\text{D}z$ is the $Q$-function.
\\

\subsection{Derivation Details}
We dedicate this subsection to the derivation of SE. This SE, as a characterization to the average MSE behavior, is obtained by averaging out all randomness from the algorithm's variances, i.e.,  $\overline{\tilde{v}^{(\ell)}}(t)$ and $\overline{\hat{v}^{(\ell)}}(t+1)$ in Algo. \ref{alg:SV-MLGAMP}.
The randomness is resulted from the input, the weighting matrices, and/or the (non-linear) activation, and the idea to average them out from the variances is motivated by the exact MMSE estimation, where an average of the variance from the target posterior equals exactly the MSE of the estimation (while a mean of the same posterior equals the output of the estimator). 
Before introducing the average, we first follow the convention of AMP \cite{bayati2011dynamics} and GAMP \cite{Rangan-arxiv10-GAMP} to made an assumption on \emph{empirical convergence} of r.v.s involved.
We say a function $\phi:\mathbb{R}^{n}\mapsto \mathbb{R}$ is \emph{pseudo-Lipschitz continuous} of order $k>1$, if there exists a constant $C>0$ such that, for any $\bs{x},\bs{y}\in \mathbb{R}^n$,
$
\|\phi(\bs{x})-\phi(\bs{y})\|\leq C(1+\|\bs{x}\|^{k-1}+\|\bs{y}\|^{k-1})\|\bs{x}-\bs{y}\|
.
$ 
Given a large sequence set $\bs{x}=\left\{x_n\right\}_{n=1}^N$, we say that, as $N\to \infty$, the component of $\bs{x}$ \emph{converges empirically} with bounded moment of order $k$ to an r.v. $X$ on $\mathbb{R}$, if
$
\lim_{N\to \infty}\frac{1}{N}\sum_{n=1}^N\phi(x_n)=\mathbb{E}\{\phi(X)\}<\infty
,
$ 
for all pseudo-Lipschitz continuous function $\phi$ of order $k$;
and this converged case is denoted by $x_n \doteq X$.
Going back to Algorithm \ref{alg:SV-MLGAMP}, we now assume that the variables $(R_{i}^{(\ell)},Z_{a}^{(\ell)})$ and $(x_i^{(\ell)}, z_a^{(\ell)})$ converge empirically to some scalars, i.e.,
\begin{align*}
\lim_{\left\{N_{\ell}\right\}\to \infty} (R_i^{(\ell)},Z_a^{(\ell)},x_i^{(\ell)},z_a^{(\ell)})\doteq(R^{(\ell)},Z^{(\ell)},x^{(\ell)},z^{(\ell)})
,
\end{align*}
and thus, ($\tilde{v}_a^{(\ell)},\hat{v}_i^{(\ell)}$) in Algorithm~\ref{alg:SV-MLGAMP} also converge empirically
\begin{align}
&\!\!\!
\overline{\tilde{v}^{(\ell)}}
\triangleq
    \frac{1}{N_{\ell+1}} \!\! \sum_{a=1}^{N_{\ell+1}}\tilde{v}_a^{(\ell)}
    \! \doteq \!
    \mathbb{E}_{R^{(\ell+1)},Z^{(\ell)}} [\tilde{v}^{(\ell)}]
    ,\,
    (\ell > 1)
\label{eq:se1}
\\
&\!\!\!
\overline{\hat{v}^{(\ell)}}
\triangleq
    \frac{1}{N_{\ell}}\sum_{i=1}^{N_{\ell}}\hat{v}_i^{(\ell)}
    \! \doteq \!
    \mathbb{E}_{R^{(\ell)},Z^{(\ell-1)}} [\hat{v}^{(\ell)}]
    ,\;
    (\ell< L)
\label{eq:se2}
\end{align}
In what follows, we evaluate the scalars above in two differnt steps and get the SE of desire.

\begin{figure}[!t]
\centering
\includegraphics[width=0.48\textwidth]{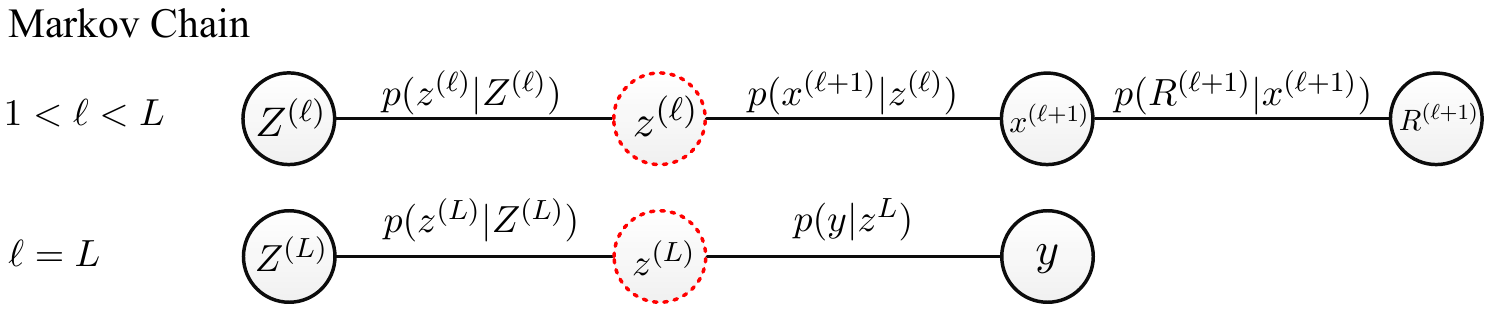}
\caption{Markov chains for the backward passing: $Z^{(\ell)} \to z^{(\ell)} \to x^{(\ell+1)} \to R^{(\ell+1)}$ and $Z^{(L)} \to z^{(L)} \to y$
}
\label{fig:Markov_Chain1}
\end{figure}

\noindent\underline{\textbf{Step 1:}} Evaluate $\overline{\tilde{v}^{(\ell)}}$ in (\ref{eq:se1}) for backward passing ($\ell<L$).
\begin{align}
\overline{\tilde{v}^{(\ell)}}
&=\mathbb{E}_{R^{(\ell+1)},Z^{(\ell)}}[\tilde{v}^{(\ell)}]\\
&=\mathbb{E}_{R^{(\ell+1)},Z^{(\ell)}}\left[\mathbb{E} [|z^{(\ell)}|^2]-|\mathbb{E}[z^{(\ell)}]|^2\right]
\label{BA4}
\end{align}
where the expectation is taken over $\mathcal{P}(R^{(\ell+1)}, Z^{(\ell)})$, while $\tilde{v}^{(\ell)}$ is the variance of a posterior distribution
\begin{align*}
\hat{p}(z_a^{(\ell)}|\bs{y})
\!\!=\!\!
\frac{\int \mathcal{N}_{x^{(\ell+1)}|z^{(\ell)}}(R^{(\ell+1)},\! \Sigma^{(\ell+1)},\! Z^{(\ell)},\! V^{(\ell)})\dd x^{(\ell+1)}}{\int \mathcal{N}_{x^{(\ell+1)}|z^{(\ell)}}(R^{(\ell+1)},\! \Sigma^{(\ell+1)},\! Z^{(\ell)},\! V^{(\ell)})\dd x^{(\ell+1)}\dd z^{(\ell)}}
\end{align*}
in which
$\mathcal{N}_{x|z}(a,A,b,B) \triangleq \mathcal{P}(x|z)\mathcal{N}(x|a,A)\mathcal{N}(z|b,B)$, as we mentioned before.
Interestingly, this posterior could be interpreted as a one-step Markov chain depicted in Fig~\ref{fig:Markov_Chain1}. Given the illustration, we express the joint p.d.f. as
\begin{align*}
&\mathcal{P}(R^{(\ell+1)},Z^{(\ell)})
=\int \mathcal{P}(R^{(\ell+1)},z^{(\ell)},Z^{(\ell)})\dd z^{(\ell)}
= \int \dd z^{(\ell)}
\\
& \mathcal{P}(Z^{(\ell)})\mathcal{P}(z^{(\ell)}|Z^{(\ell)})\mathcal{P}(x^{(\ell+1)}|z^{(\ell)})\mathcal{P}(R^{(\ell+1)}|x^{(\ell+1)})\dd x^{(\ell+1)}
\end{align*}
where
$\mathcal{P}(R^{(\ell+1)}|x^{(\ell+1)})=\mathcal{N}(x^{(\ell+1)}|R^{(\ell+1)},\Sigma^{(\ell+1)})$,
$\mathcal{P}(z^{(\ell)}|Z^{(\ell)}) \!=\! \mathcal{N}(z^{(\ell)}|Z^{(\ell)}, \! V^{(\ell)})$, and $\mathcal{P}(Z^{(\ell)})$ is solved from
\begin{align}
\int \mathcal{P}(Z^{(\ell)})\mathcal{P}(z^{(\ell)}|Z^{(\ell)})\dd Z^{(\ell)}=\mathcal{P}(z^{(\ell)})
\label{BA3}
\end{align}
To solve the equation, we notice for its r.h.s. that in large system limits, where all weighting matrices are drawn from some i.i.d. zero-mean populations with $N_{\ell} \to \infty$ and $N_{\ell+1}/N_{\ell} \to \alpha_{\alpha}$, the r.v. $z^{(\ell)}$ tends to be Gaussian distributed with mean zero and variance $T_z^{(\ell)}$, according to the central limit theorem. Hence, we have $\mathcal{P}(z^{(\ell)})=\mathcal{N}(z^{(\ell)}|0,T_z^{(\ell)})$, where
\begin{align*}
&
T_z^{(\ell)}
\triangleq
\mathbb{E}_{\bs{H}^{(\ell)},\bs{x}^{(\ell)}} [|z^{(\ell)}|^2]
=
    \mathbb{E}_{\bs{x}^{(\ell)}} \{
    (\sum_{i=1}h_{ai}x_{i}^{(\ell)}) (\sum_{j=1}h_{aj}x_{j}^{(\ell)})^{*}
    \}
\\
&   \quad\quad
    =\sum_{i=1}^{N_{\ell}}\mathbb{E}_{\bs{H}^{(\ell)}}\left\{|h_{ai}^{(\ell)}|^2\right\}\mathbb{E}_{\bs{x}^{(\ell)}}\left[|x_i^{(\ell)}|^2\right]
    ={T_x^{(\ell)}}/{\alpha_{\ell}}
\\
&T_x^{(\ell)}
=
    \!\! \int \!\! |x^{(\ell)}|^2 \mathcal{P}(x^{(\ell)}|z^{(\ell-1)})
    \mathcal{N}(z^{(\ell-1)}|0,T_z^{(\ell-1)})\dd z^{(\ell-1)}\dd x^{(\ell)}
\end{align*}
Particularly, at $\ell=1$, $T_x^{(1)} =\sigma_X^2$. Now going back to the solution of the equation (\ref{BA3}). Given that  $\mathcal{P}(z^{(\ell)}|Z^{(\ell)})$ and $\mathcal{P}(z^{(\ell)})$ are both Gaussian, it is easy to show (by de-convolution) that the function $\mathcal{P}(Z^{(\ell)})$ is also Gaussian, and could be expressed as $\mathcal{P}(Z^{(\ell)})=\mathcal{N}(Z^{(\ell)}|0,T_z^{(\ell)}-V^{(\ell)})$. Putting all these together, we get the joint p.d.f. rewritten as
\begin{align}
&\mathcal{P}
(R^{(\ell+1)},Z^{(\ell)})
=\mathcal{N}(Z^{(\ell)}|0,T_z^{(\ell)}-V^{(\ell)}) \times
\nonumber\\& \;
\int \! \mathcal{N}_{x^{(\ell+1)}|z^{(\ell)}}(R^{(\ell+1)},\Sigma^{(\ell+1)},Z^{(\ell)},V^{(\ell)})\dd z^{(\ell)}\dd x^{(\ell+1)}
\label{joint_RZ}
\end{align}
Next, we substitute the joint p.d.f. back into (\ref{BA4}) and perform some algebraic manipulations to get the followings
\begin{align}
\overline{\tilde{v}^{(\ell)}}
&=
    T_z^{(\ell)}-q^{(\ell)}
\label{CA5}
\\
q^{(\ell)}
& =
    \frac{\left|\int
    z^{(\ell)}\mathcal{N}_{x^{(\ell+1)}|z^{(\ell)}} (\cdot)
    \dd x^{(\ell+1)} \dd z^{(\ell)}\right|^2}
    {\int \mathcal{N}_{x^{(\ell+1)}|z^{(\ell)}} (\cdot) \dd x^{(\ell+1)} \dd z^{(\ell)} }\text{D}\xi\dd \zeta
\\
(\cdot)
&=(\zeta,\Sigma^{(\ell+1)},\sqrt{T_z^{(\ell)}-V^{(\ell)}}\xi,V^{(\ell)})
\end{align}
with
$\text{D}\xi\triangleq \mathcal{N}(\xi|0,1)\dd \xi$.
In particular, at $\ell=L$, we have
\begin{align}
\overline{\tilde{v}^{(L)}}
&=
    T_z^{(L)}-q^{(L)}
\label{CA4}
\\
q^{(L)}
&=
    \int \frac{\left|\int z^{(L)}\mathcal{P}(y|z^{(L)})\mathcal{N}(z^{(L)}|\cdot)\dd z^{(L)}\right|^2}
{\int \mathcal{P}(y|z^{(L)})\mathcal{N}(z^{(L)}|\cdot)\dd z^{(L)}}\text{D}\xi\dd y
\end{align}
where $\mathcal{N}(z^{(L)}|\cdot)$ denotes $\mathcal{N}(z^{(L)}|\sqrt{T_z^{(L)}-V^{(L)}}\xi,V^{(L)})$.

\begin{figure}[!t]
\centering
\includegraphics[width=0.48\textwidth]{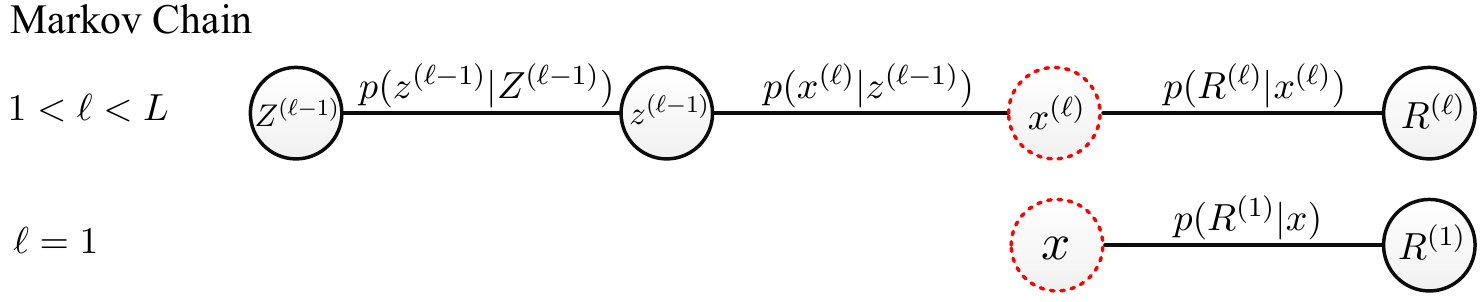}
\caption{Markov chains for the forward passing: $Z^{(\ell-1)} \to z^{(\ell-1)} \to x^{(\ell)} \to R^{(\ell)}$ and $x\to R^{(1)}$
}
\label{fig:Markov_Chain2}
\end{figure}

\noindent\underline{\textbf{Step 2:}} Evaluate $\overline{\hat{v}^{(\ell)}}$ in (\ref{eq:se2}) for forward passing ($\ell>1$).
\begin{align}
\overline{\hat{v}^{(\ell)}}
&=\mathbb{E}_{R^{(\ell)},Z^{(\ell-1)}}\left[\hat{v}^{(\ell)}\right]\\
&=\mathbb{E}_{R^{(\ell)},Z^{(\ell-1)}}\left[\mathbb{E}\left[|x^{(\ell)}|^2\right]-|\mathbb{E}[x^{(\ell)}]|^2\right]
\label{BA5}
\end{align}
where $\hat{v}^{(\ell)}$ is the variance of an approximate posterior
\begin{align*}
\hat{p}(x_i^{(\ell)}|\bs{y})
\!\!=\!\!
\frac{\int \mathcal{N}_{x^{(\ell)}|z^{(\ell-1)}}(R^{(\ell)},\Sigma^{(\ell)},Z^{(\ell-1)},V^{(\ell-1)})\dd z^{(\ell-1)}}{\int \mathcal{N}_{x^{(\ell)}|z^{(\ell-1)}}(R^{(\ell)},\Sigma^{(\ell)},Z^{(\ell-1)},V^{(\ell-1)})\dd z^{(\ell-1)}\dd x^{(\ell)}}
\end{align*}
which could also be interpreted via a Markov chain as shown in Fig.~\ref{fig:Markov_Chain2}.
We then evaluate (\ref{BA5}) in a way similar to the last step and get
\begin{align}
\overline{\hat{v}^{(\ell)}}
&=
T_x^{(\ell)}
-d^{(\ell)}
\label{CA6}
\\
d^{(\ell)}
&=
\int
\frac{
        \left|
        \int x^{(\ell)} \mathcal{N}_{x^{(\ell)}|z^{(\ell-1)}}(\cdot)
        \, \dd z^{(\ell-1)}\dd x^{(\ell)}
        \right|^2
    }{
        \int \mathcal{N}_{x^{(\ell)}|z^{(\ell-1)}}(\cdot)
        \, \dd z^{(\ell-1)}\dd x^{(\ell)}
    }
    \text{D}\xi\dd \zeta
\\
(\cdot)
& =
(\zeta,\Sigma^{(\ell)},\sqrt{T_z^{(\ell-1)}-V^{(\ell-1)}}\xi,V^{(\ell-1)})
\end{align}
and if $\ell=1$, we further have
\begin{align}
\overline{\hat{v}^{(1)}}
&=T_x^{(1)}-d^{(1)}
=\varepsilon(\Sigma^{(1)})
\\
T_x^{(1)}
&=\int |x|^2 \mathcal{P}_X(x)\dd x=\sigma_X^2
\\
d^{(1)}
&=\int \frac{\left|\int x \mathcal{P}_X(x)\mathcal{N}(x|\zeta,\Sigma^{(1)})\dd x\right|^2}{\int \mathcal{P}_X(x)\mathcal{N}(x|\zeta,\Sigma^{(1)})\dd x}
\dd \zeta
\end{align}

So far, we have obtained the explicit expressions for the (empirically converged) variables, $\overline{\tilde{v}^{(\ell)}}$,
$\Sigma^{(\ell)}$,
$\overline{\hat{v}}^{(\ell)}$,
$V^{(\ell)}$,
$q^{(\ell)}$,
$d^{(\ell)}$,
$T_x^{(\ell)}$,
and
$T_z^{(\ell)}$).
By substituting them back into Algo. \ref{alg:SV-MLGAMP}, we finally get the SE of desire.

\section{Numerical Examples} \label{sec:simulations}
In this section, we verify the performance of the ML-GAMP (Algorithm \ref{alg:CML-GAMP}, the accuracy of the SE characterization (Algorithm \ref{alg:CML-GAMP}), and the agreement with the replica fixed point equations (Claim 1 in Part I). A comparison to the existing ML-GAMP \cite{manoel2017multi} is also provided, confirming our findings on the convergence issue.

\subsection{Parameter Setup}
\begin{table}[!t]
\centering
\caption{Parameter setting of the numerical examples}
\label{Table_a}
\begin{tabular}{|l|c|c|l|l|}

  \hline
               & Input & Layers & Each Layer's Model & Weighting\\
  \hline
  Ex. \#1 & QPSK  & 2   &  AWGN; AWGN+ADC & $N_{\ell}=N_{\ell-1}$ \\
  \hline
  Ex. \#2 & QPSK  & 2$\sim$4   & AWGN; AWGN; ...  & 
  $N_{\ell}=2N_{\ell-1}$ \\
  \hline
  Ex. \#3 & QPSK  & 2   & AWGN; AWGN+ADC & $N_{\ell}=2N_{\ell-1}$\\
  \hline
\end{tabular}
\end{table}

Table \ref{Table_a} summarizes the settings of three examples considered here. An analog-to-digital convertor (ADC) of uniform quantization is adopted to exemplify our applicability to non-linear activation, and it is appended to an AWGN channel, resulting in a signal model as below (for a certain layer):
\begin{eqnarray}
\bs{z}
& = &
\bs{H} \bs{x}
\\
\tilde{\bs{y}}
& = &
    \bs{z} + \bs{n}
\\
\bs{y}
& = &
    Q_c(\tilde{\bs{y}})
\end{eqnarray}
where $Q_c(\cdot)$ is the complex-valued quantization that reads
\begin{align*}
y_m=Q_c(\tilde{y}_m)=Q(\Re(\tilde{y}_m))+\mathbb{J}Q(\Im(\tilde{y}_m))
,
\quad m=1,\cdots,M
\end{align*}
with $Q(\cdot):\mathbb{R}\mapsto \mathcal{R}_B$, and $\mathcal{R}_B$ being the set of $b$-bits defined as:
$
\mathcal{R}_B\triangleq
\{\left(-\frac{1}{2}+b\right)\triangle;\ b=-\frac{2^B}{2}+1,\cdots,\frac{2^B}{2}\},
$
where $\triangle$ denotes the uniform quantization step.
For uniform quantization, the output of ADC is assigned a value $y$ when the input $\tilde{y}$ falls within the  range of $(q^{\text{low}}(y),q^{\text{up}}(y)]$, where
\begin{align}
q^{\text{low}}(y_m)&=\left\{
\begin{array}{cl}
y_m-\frac{\triangle}{2}, &\text{if} \ y_m\geq -\left(\frac{2^B}{2}-1\right)\triangle, \\
-\infty, &\text{otherwise}.
\end{array}
\right.\\
q^{\text{up}}(y_m)&=\left\{
\begin{array}{cl}
y_m+\frac{\triangle}{2}, &\text{if} \ y_m\leq \left(\frac{2^B}{2}-1\right)\triangle, \\
\infty, &\text{otherwise}.
\end{array}
\right.
\end{align}
Adding ADC, the overall transitional probability from a complex $z$ to a complex $y$ then becomes
\begin{align*}
\mathcal{P}(y|z)=\Psi(\Re(y)|\Re(z),\frac{\sigma_2^2}{2}) \cdot \Psi(\Im(y)|\Im(z),\frac{\sigma_2^2}{2})\cdot \mathbbm{1}_{y\in \mathcal{R}_B+\mathbb{J} \mathcal{R}_B}
\end{align*}
where $\Psi(y|z,c^2) = \Phi\left(\frac{y-z}{c}\right)$ and $\Phi(x)=\int_{-\infty}^{x}\mathcal{N}(t|0,1)\dd t$.

\subsection{Experimental Results}

\begin{figure}[!t]
\centering
\includegraphics[width=0.53\textwidth]{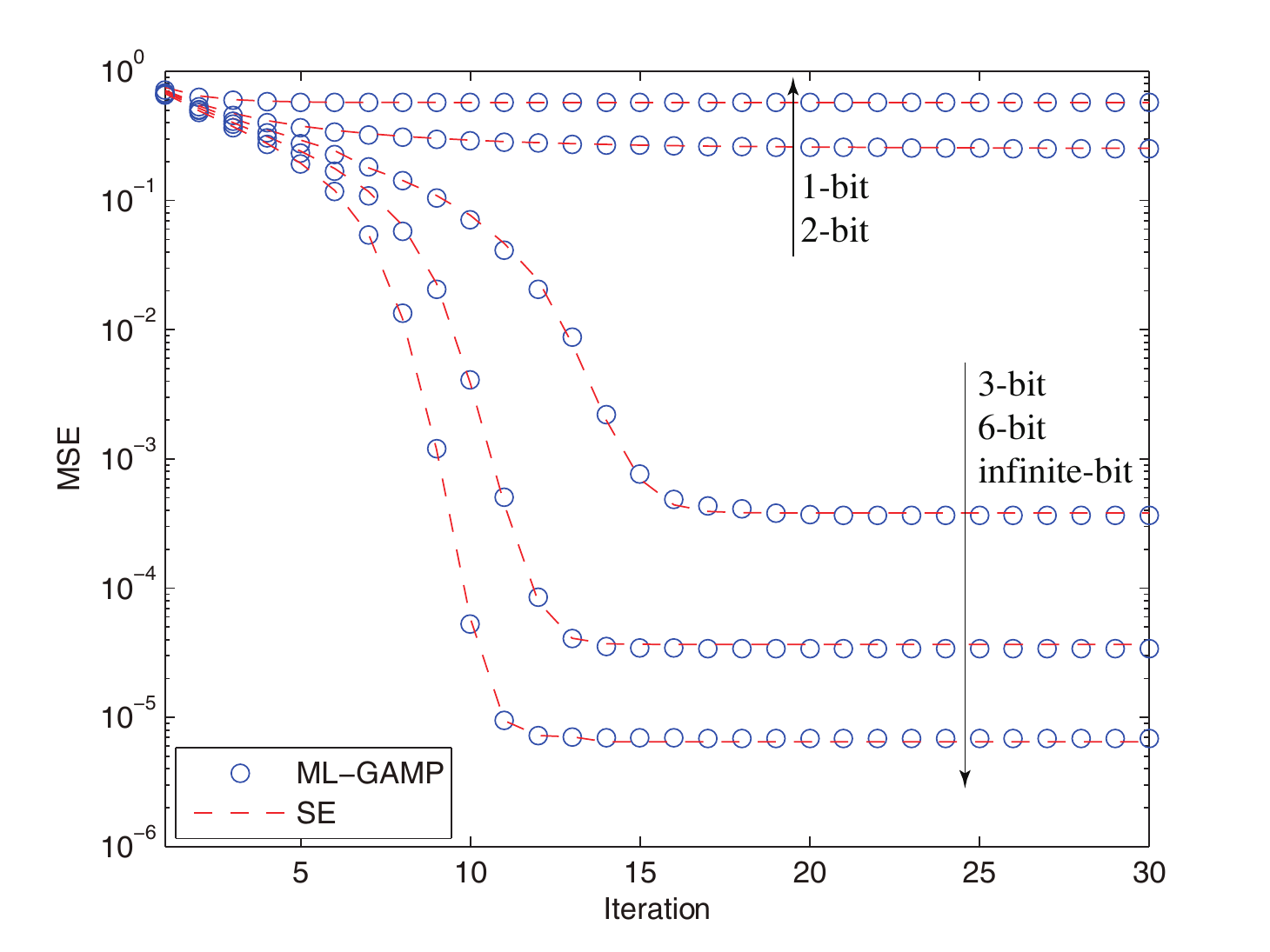}
\caption{Example 1, comparisons between Monte-Carlo results and the SE evaluation (also the replica prediction), confirming the perfect agreement.
}
\label{sim_iteration}
\end{figure}

In Example \#1, we consider a 2-layer GLM case, where the first layer is AWGN, and the second layer is AWGN+ADC. The sizes of the two transform matrices are uniformly set to be 1024, and the signal-to-noise ratios (SNRs) of the two layers are $20$dB and $15$dB, respectively.
We vary the ADC resolution, from 1-bit, 2-bit, 3-bit, and 6-bit to $\infty$-bit, to verify the agreements of the proposed algorithm's simulated MSEs with its SE.
For the AWGN+ADC cased existing in second layer, we have following notes
\footnote{
A major difficulty in particularizing the general SE into a specific AWGN+ADC case is the evaluation of $\overline{\tilde{v}^{(2)}}$, which involves non-linear activation. To handle this, we take measures similar to \cite{wen2015bayes, wang2017bayesian} and obtain
\begin{align}
\overline{\tilde{v}^{(2)}}=V^{(2)}-\frac{(V^{(2)})^2}{\sigma_2^2+V^{(2)}}\sum_{y\in \mathcal{R}_B}\int_{\xi}\frac{[\phi(\eta_1(y))-\phi(\eta_2(y))]^2}{\Phi(\eta_1(y))-\Phi(\eta_2(y))}\text{D}\xi
\end{align}
where
\begin{align*}
\eta_1(y)=\frac{q^{\text{up}}(y)-\sqrt{\frac{T_z^{(2)}-V^{(2)}}{2}}\xi}{\sqrt{\frac{V^{(2)}+\sigma_2^2}{2}}},\ \eta_2(y)=\frac{q^{\text{low}}(y)-\sqrt{\frac{T_z^{(2)}-V^{(2)}}{2}}\xi}{\sqrt{\frac{V^{(2)}+\sigma_2^2}{2}}}
\end{align*}
In addition, the variable $\overline{\hat{v}^{(1)}}$ depends on the prior. In case of a QPSK prior, $\overline{\hat{v}^{(1)}}=1-\int \tanh((\Sigma^{(1)})^{-1}+\sqrt{(\Sigma^{(1)})^{-1}}\zeta)\text{D}\zeta$. Other variables like $\overline{\tilde{v}^{(1)}}$ and $\overline{\hat{v}^{(2)}}$ could be evaluated readily via Gaussian reproduction property \cite{Rasmussen-Book06-GaussianReproduction}.
}.
By MSE, we refer to the normalized version that is obtained via
$
{\|\bs{x}-\hat{\bs{x}}\|_2^2} / {\|\bs{x}\|_2^2}
$.
Then we conclude from Fig \ref{sim_iteration}:\\
$\bullet$
The SE results (obtained via numerical evaluation) agree perfectly with the algorithm results (obtained through Monte Carlo simulations), confirming that the SE is capable of capturing the dynamics of the ML-GAMP algorithm, and also that the fixed points predicted by the replica analysis for a generic MMSE estimator could indeed be realized by an ML-GAMP estimator.
\\
$\bullet$
We required, during the theoretical analysis, the system dimension should be large enough (usually tends to infinity) so that theories like central limit theorem and large deviation theory would be true. In practice, a $1024 \times 1024$ matrix size seems good enough already. Besides, the ML-GAMP algorithm admit a fast converge speed, i.e., in all cases we have considered, it converged with less $15$ iterations.

\begin{figure}[!t]
\centering
\includegraphics[width=0.48\textwidth]{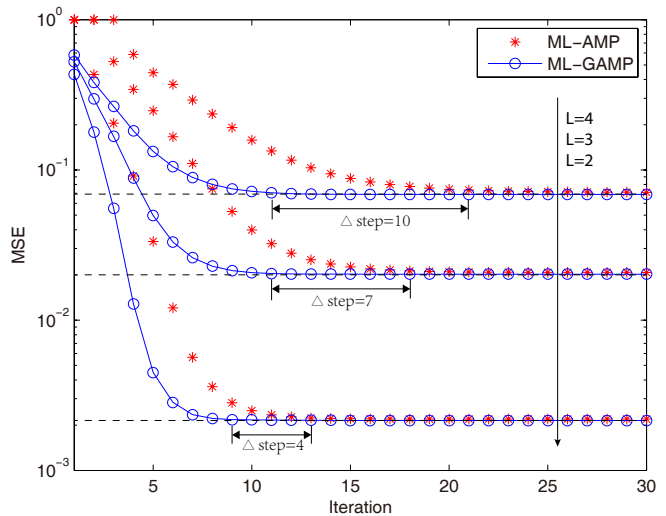}
\caption{Example 2, comparison between the proposed ML-GAMP and the existing ML-AMP \cite{manoel2017multi}, confirming ML-GAMP's faster convergence rate.
}
\label{fig:sim_MLGAMP&MLAMP}
\end{figure}

In Example \#2, we consider 3 different settings of ML-GLM, i.e., $L=2$, $L=3$, and $L=4$, all having AWGN in each layer. The SNR of each layer is fixed at $10$dB, and the matrix dimension is in a doubling manner, i.e., $N_{\ell+1}=2N_{\ell}$ for $\ell=1,\cdots,L$ with $N_1=256$. In this setting, we compare the MSE performance of the proposed ML-GAMP to that from ML-AMP \cite{manoel2017multi}. This result is given as Fig~\ref{fig:sim_MLGAMP&MLAMP}, suggesting:
\\
$\bullet$
Both ML-GAMP and ML-AMP converge to the same MSE, which (recall from last example's result) is indeed the optimal value a generic MMSE estimator would attain.
\\
$\bullet$
The difference is our ML-GAMP converges in a speed faster than the existing ML-AMP. As the number of layers increases, the superior of our algorithm becomes dominant. As we analyzed earlier, this superiority is owing to a more recent update of messages in our algorithm, and the superiority would accumulate with the increase of layer number.

\begin{figure}[!t]
\centering
\includegraphics[width=0.53\textwidth]{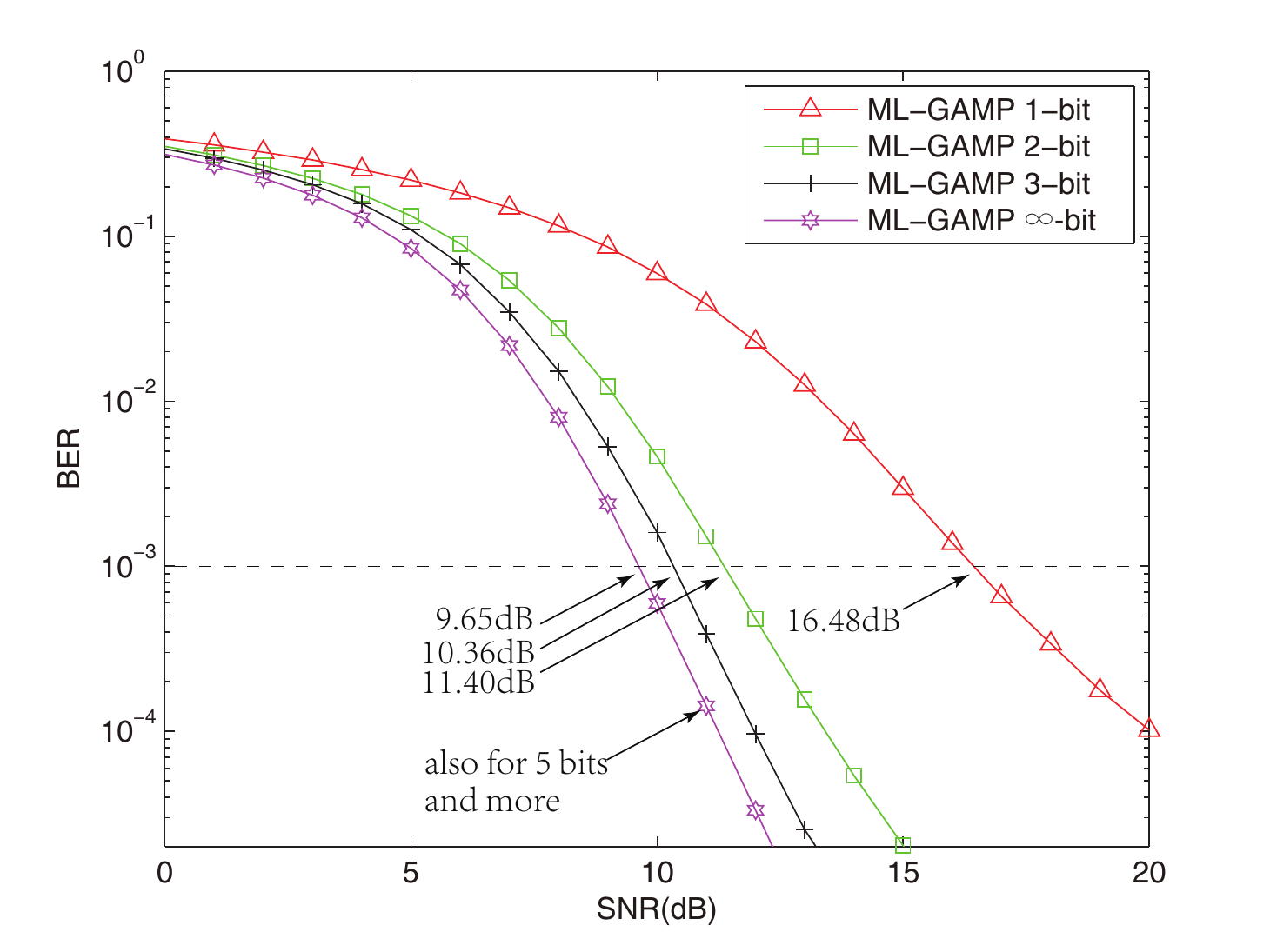}
\caption{Example 3, the impact of the ADC resolution on BER performance, revealing a diminishing marginal effect as the resolution ramps up.
}
\label{sim_SNR}
\end{figure}

In Example \#3, we consider a two-layer GLM, where the 1st layer is a $256\times 512$ AWGN setting while the 2nd is a $512 \times 1024$ AWGN+ADC setting. We vary the ADC resolution from 1-bit, 2bit, 3-bit, to $\infty$-bit to see the impact of uniform quantization. From Then, we find from Fig~\ref{sim_SNR}: \\
$\bullet$
Increasing the ADC resolution (i.e., number of quantization bits) does help in reducing the bit error rate (BER). This is in line with our intuition that the ADC quantization is a lossy processing; the more bits added, the less loss we see.
\\
$\bullet$
Increasing the resolution of the ADC, however, only provides a diminishing marginal benefit. In an extreme case as we illustrate, after $B \geq 5$, the benefit become so trivial that is not even noticeable, and this is because its BER has already hit the lower bound, i.e. AWGN without ADC.

\section{Conclusions} \label{sec:conclusions}
As Part II of the two-part work, this paper proposed a new estimator, the ML-GAMP, to work around the implementation difficulty of an exact MMSE estimator in high-dimensional ML-GLM. The new estimator was derived using a method adopted in our previous paper \cite{zou2018concise}, which blended an moment-matching projection into the LBP framework.   Comparing to existing results, the new estimator provided a better tradeoff between efficiency and complexity. On one hand, it converged in a speed much faster than the ML-AMP \cite{manoel2017multi}, while on the other hand, it avoided the use of an expansive SVD operation that was required by the ML-VAMP \cite{Fletcher-arxiv17-ML-VAMP}. Further analysis in the new estimator also revealed that its asymptotic MSE behavior could be fully characterized by a set of iterating equations in scalar form, called SE. This SE shared exactly the same fixed points as the exact MMSE estimator whose fixed points were obtained in Part I via replica analysis.
Since the exact MMSE estimator is Bayes-optimal in an MSE sense, such an coincidence in the fixed points suggested that the  proposed estimator (if converged) could offer an estimate that is MSE optimal.

\section{Acknowledgement}
This work was supported in part by the Scientific Research Fund of Guangzhou 201904010297, Natural Scientific Fund of Guangdong 2016A030313705, Special Fund for Applied Science and Technology of Guangdong 2015B010129001, and Natural Scientific Fund of Guangxi 2018GXNSFDA281013.

\ifCLASSOPTIONcaptionsoff
  \newpage
\fi


\end{document}